\documentclass[12pt,a4paper]{article}
\usepackage{amssymb}
\usepackage{amsfonts}
\usepackage[onehalfspacing]{setspace}
\usepackage{geometry}

\newtheorem{claim}{Claim}

\newtheorem{definition}{Definition}

\newtheorem{proposition}{Proposition}
\newtheorem{remark}{Remark}

\input{tcilatex}
\begin{document}

\title{News Media as Suppliers of Narratives \\
(and Information)\thanks{%
Spiegler acknowledges financial support from Leverhulme Trust grant
RPG-2023-120. We thank Alex Clyde, Daniel Gottlieb, Shira Dvir-Gvirtsman,
and the seminar audience at the Hebrew University for helpful comments.}}
\author{Kfir Eliaz and Ran Spiegler\thanks{%
Eliaz: School of Economics, Tel-Aviv University and David Eccles School of
Business, University of Utah. E-mail: kfire@tauex.tau.ac.il. Spiegler:
School of Economics, Tel-Aviv University and Economics Dept., University
College London. E-mail: rani@tauex.tau.ac.il.}}
\maketitle

\begin{abstract}
We present a model of news media that shape consumer beliefs by providing
information (signals about an exogenous state) and narratives (models of
what determines outcomes). To amplify consumers' engagement, media maximize
consumers' anticipatory utility. Focusing on a class of separable consumer
preferences, we show that a monopolistic media platform facing homogenous
consumers provides a false \textquotedblleft empowering\textquotedblright\
narrative coupled with an optimistically biased signal. Consumer
heterogeneity gives rise to a novel menu-design problem due to a
\textquotedblleft data externality\textquotedblright\ among consumers. The
optimal menu features multiple narratives and creates polarized beliefs.
These effects also arise in a competitive media market model.\bigskip
\bigskip \bigskip \bigskip \pagebreak
\end{abstract}

\section{Introduction}

Standard models of news media regard them as suppliers of information,
providing noisy signals of an underlying state of Nature. A complementary
view, which is absent from standard models, is that news media are a vehicle
for spreading \textit{narratives}, as reflected in the following quotes by
prominent journalists:\medskip

\begin{quote}
\textit{\textquotedblleft It's all storytelling, you know. That's what
journalism is all about.\textquotedblright } (Tom Brokaw)\medskip

\textit{\textquotedblleft We're supposed to be tellers of tales as well as
purveyors of facts. When we don't live up to that responsibility, we don't
get read.\textquotedblright } (William Blundell)\medskip
\end{quote}

\textquotedblleft Stories\textquotedblright\ or \textquotedblleft
narratives\textquotedblright\ are of course loaded terms with rich meanings
in the context of news reporting. We conceive of narratives as models or
frames that condition media consumers' thinking about the significance of
reported information. For example, while many exogenous variables can be
reported, the media often selects only some of them as relevant for outcomes
of interest and therefore worthy of reporting. Another example is the
shaping of popular perceptions about the role of personal agency and
external factors in life outcomes. In particular, when reporting about
discrimination, the media can peddle a narrative that focuses on the role of
personal effort in achieving material success. Alternatively, it can offer a
narrative that attributes economic outcomes solely to discrimination; or a
complex narrative that incorporates both factors.\footnote{%
This example echoes Glenn Loury's (2020) distinction between
\textquotedblleft development\textquotedblright\ and \textquotedblleft
bias\textquotedblright\ narratives.}

A related example involves narratives about financial investments. A popular
narrative aimed at retail investors is that the value of their portfolio
depends on how they manage it (how active are they? how do they allocate the
portfolio between types of risk?). Such a narrative is misleading because it
neglects the possibility that market fundamentals (interest rates, business
sectors' risk or growth potential) are already reflected in security prices
and thus affect investment returns. Thus, when financial-news media report
about market fundamentals (e.g., that certain sectors are experiencing
growth), the simplistic narrative may lead media consumers to draw wrong
conclusions from this information (e.g., that they should invest in a growth
sector).

This paper presents a model of news media (in a broad sense that includes
content platforms) that is based on a fusion of the two views: The media
provides information about exogenous states as well as a narrative, which is
a model of the determination of outcomes as a function of states and
actions. Media consumers use the narrative to interpret statistical
regularities, forming beliefs about the mapping from states and actions to
outcomes. A false narrative is a misspecified model, which can therefore
induce distorted beliefs.

The fusion of the information-based and narrative-based views enables us to
offer a new model of \textit{media bias}. There is a common intuition that
this phenomenon is driven in large part by consumer demand (Gentzkow and
Shapiro (2010) back this intuition with empirical evidence). Yet, the
standard model of consumer behavior assumes that demand for information is
purely instrumental. Expected-utility maximizers weakly prefer more
informative signals. Therefore, unless there are frictions on the supply
side that prevent media from providing complete and objective information,
the market will provide it. Even if consumers have heterogeneous
preferences, they all want more informative news.

Studies across several disciplines (psychology, political science and
communication) have shown evidence that consumer demand for news media
reflects non-instrumental attitudes to beliefs (e.g., Hart et al. (2009),
Van der Meer et al. (2020), Taber and Lodge (2006)). These findings have
inspired models of media bias in which beliefs enter directly into
consumers' utility function (see Prat and Str\"{o}mberg (2013) and Gentzkow
et al. (2015) for surveys). We propose a model of demand for news in which
consumers have \textit{intrinsic preferences over their posterior beliefs},
which they arrive at via Bayesian updating of prior beliefs shaped by the
narrative they adopt. To our knowledge, this integration of non-instrumental
and Bayesian aspects of consumer beliefs is new to the literature on media
bias. We discuss the relation to relevant literature in Section 6.

One can distinguish between two types of intrinsic preferences over
posterior beliefs. \textit{Retrospective} preferences rank beliefs about
past outcomes --- e.g., wanting to believe that one is the just side in a
dispute (see Chopra et al. (2023)). \textit{Prospective} preferences rank
beliefs about future outcomes, taking into account private or public actions
in response to beliefs. We focus on the latter type: Consumers approach news
media with the desire to maximize their \textit{anticipatory utility} ---
i.e., the expected indirect utility from their posterior beliefs. According
to this view, demand for news is driven by consumers' pursuit of optimistic
posterior beliefs. By \textquotedblleft optimism\textquotedblright , we do
not necessarily mean that the media paint a rosy picture of reality, but
rather that it gives \textit{hope} that if the right decisions are made,
future prospects will be good. Real-life manifestations of this idea include
patriotic coverage of international conflicts or sporting events that
amplify the prospect of victory;\footnote{%
In discussing the popularity of patriotic coverage of the war in Afganistan
and Iraq, a New York Times story (Ruthenberg (2003)) quotes MSNBC's
president Erik Sorenson:\ \textquotedblleft After Sept. 11 the country wants
more optimism and benefit of the doubt...It's about being positive as
opposed to being negative.\textquotedblright} business-news media conveying
the impression that retail financial investors can beat the market; and
reports of police brutality or climate change which send a message that
inflates the ability of policy reforms (\textquotedblleft defunding the
police\textquotedblright , switching to green energy) to improve social
welfare.

However, under the conventional assumption that news media only supply
information, prospective preferences over (Bayesian) posterior beliefs
cannot give rise to media bias. The reason is that expected anticipatory
utility is the upper envelope of linear functions of posterior beliefs
(hence convex in these beliefs). As a result, when the media caters to
consumer demand by offering a signal function that maximizes consumers'
anticipatory utility, it will weakly prefer full information provision.
Thus, even when we allow for prospective non-instrumental demand for
information, the standard view of the media as information providers cannot
generate media bias.

This is where our view of media as joint providers of narratives and
information enters. We show that this more comprehensive approach provides a
non-trivial model of media bias, such that distortion of the truth consists
of biased/inaccurate reports \textit{together with} false narratives.
Moreover, there is a synergy between these two instruments: They complement
each other in producing the hopeful beliefs that consumers seek.\bigskip

\noindent \textit{Overview of model and results}

\noindent In the basic version of our model, a representative consumer takes
an action after observing a signal about a state of Nature. There is an
objective stochastic mapping from states and actions to outcomes. The
consumer is endowed with a vNM utility function which is separable in the
action. A monopolistic media outlet commits ex-ante to a \textquotedblleft
media strategy\textquotedblright , which consists of: $(i)$ a Blackwell
experiment (namely, a stochastic mapping from states to signals), and $(ii)$
a narrative, which selects a subset of the outcome's true causes.

There are four feasible narratives. The true narrative acknowledges both
states and actions as causes. The \textquotedblleft
empowering\textquotedblright\ narrative postulates that actions are the sole
cause of outcomes (in terms of the discrimination example above, this
narrative says that only personal effort determines economic outcomes, thus
suppressing the role of discrimination). The \textquotedblleft
fatalistic\textquotedblright\ narrative postulates that only the state
matters for the outcome (e.g., it says that only the external force of
discrimination determines outcomes, thus suppressing the role of personal
agency). Finally, the \textquotedblleft denial\textquotedblright\ narrative
removes both the action and the state as causes, thus implicitly attributing
outcomes to unspecified other factors.

Given an empirical long-run distribution over states, actions and outcomes,
a narrative produces a subjective (and possibly distorted) belief by
\textquotedblleft fitting\textquotedblright\ the narrative to this
distribution. For example, the empowering narrative interprets the empirical
correlation between actions and outcomes as a causal quantity --- i.e., it
attributes the variation in outcomes entirely to variation in actions. Once
the consumer adopts a narrative, his strategy (a stochastic mapping from
signals to actions) prescribes actions that maximize expected utility with
respect to the narrative-induced belief. In equilibrium, this strategy is
consistent with the empirical long-run distribution. The need for an
equilibrium definition of consumer response to a given narrative is typical
of models of decision making under misspecified models (e.g., Esponda and
Pouzo (2016), Spiegler (2016), Eliaz and Spiegler (2020)). The reason is
that changes in long-run behavior can lead to changes in the consumer's
perceived mapping from actions to outcomes.

The media's problem is to find a strategy and an equilibrium (induced by the
strategy) that maximize the consumer's ex-ante expected anticipatory
utility. The rationale for maximizing anticipatory utility is that a
consumer's engagement with the media increases with the amount of optimism
he can derive from its consumption. The better the media performs in
generating optimistic beliefs, the higher the demand for it. A crucial
feature of the problem is that the media takes into account equilibrium
effects when designing its strategy. This is similar in spirit to
information-design problems (e.g., Bergemann and Morris (2019)). However, in
standard models equilibrium effects arise in multi-agent settings with
payoff externalities. In contrast, in our model equilibrium effects arise
because of misspecified beliefs induced by false narratives.

This account of news media raises a number of questions: Will the media
provide accurate, unbiased information? If not, what is the structure of
media inaccuracy/bias, and which narratives will it peddle? Our analysis of
the baseline model in Section 3 addresses these questions. We begin with a
full characterization of the optimal media strategy in a specification of
our model --- inspired by the above discrimination and financial-investment
scenarios --- which serves as a running example in our paper. The optimal
strategy consists of the empowering narrative and a signal with an
optimistic bias (i.e., always correctly reporting good news and sometimes
misrepresenting bad news). The magnitude of the bias is tailored to consumer
preferences.

We then show that this combination is a robust feature: For any
action-separable utility function, if a media strategy outperforms the
true-narrative, perfect-information benchmark, then it must involve the
empowering narrative. Also, it must provide information that induces
different behavior from the benchmark (as long as the benchmark leads to
state-contingent actions). Thus, \textit{there is a synergy between false
narratives and biased information, which is essential for the media's
mission to maximize consumers' anticipatory utility}. This result also
demonstrates the value of our model in making specific predictions about the
structure of media narratives and media bias.

Section 4 introduces preference heterogeneity among consumers in the context
of our running example. This naturally calls for a model in which consumers
can choose between multiple narrative-information combinations. We now
envisage our monopolistic media provider as a gatekeeper or a platform that
restricts the entry of these combinations. Effectively, this means that the
media chooses a menu of media strategies, aiming to maximize aggregate
anticipatory utility.\footnote{%
For tractability, we restrict media strategies to report good news in the
good state, which is an endogenous feature of the optimal strategy in the
baseline model.} From the menu, each consumer chooses the
narrative-information pair that maximizes his own anticipatory utility,
evaluated according to the equilibrium joint distribution over states,
actions and outcomes that arises from the aggregate behavior of \textit{all}
consumer types.

At first glance, it may appear that incentive-compatibility should be moot
in this model, because media and consumers have a common objective. However,
this is not the case because of the \textquotedblleft \textit{data
externality}\textquotedblright\ that exists among consumer types. Although
they are separate individuals with idiosyncratic preferences, they all rely
on the same aggregate data to form beliefs given the narratives they adopt.
Consequently, changes in the behavior of one segment of the consumer
population can change how another segment evaluates narrative-information
pairs. Dealing with this externality in the context of a menu design problem
is a methodological\ novelty of our paper, and one of our motivations for
introducing heterogeneity in the first place.

The data externality turns out to have a significant effect on the optimal
menu, compared with the representative-consumer case. The contrast is
particularly stark when consumer types are uniformly distributed. Instead of
biased, partially informative signals that are finely tailored to consumer
types, now none of the consumers receive $any$ information. Nevertheless,
the population is split into two \textquotedblleft camps\textquotedblright\
with starkly different beliefs, driven by the different narratives they
adopt. One segment opts for the empowering narrative and takes one constant
action, while the other segment opts for the denial narrative and always
takes the opposite action. Thus, our model shows how a heterogeneous
population of consumers trying to make sense of the same aggregate data can
end up holding highly polarized beliefs based on no information, simply
because of the narrative-peddling aspect of media strategy.

We then explore the role of market structure by examining a
\textquotedblleft perfect competition\textquotedblright\ version of the
heterogeneous-consumers model. Each media provider is \textquotedblleft
small\textquotedblright\ in the sense that it takes the joint distribution
over states, actions and outcomes as given, without taking into account how
its media strategy affects this distribution via its effect on consumers'
beliefs and actions. A \textquotedblleft competitive
equilibrium\textquotedblright\ is a profile of media strategies, one for
each consumer type, such that: $(i)$ the strategy associated with a type
maximizes his anticipatory utility given the aggregate distribution; and $%
(ii)$ this distribution arises from each type best-replying to the belief
induced by the media strategy he adopts. We show that in the essentially
unique equilibrium, only the true and fatalistic narratives prevail, where
the former narrative is coupled with complete information. When consumer
types are uniformly distributed, the segment of consumers who act on
informative signals is larger than in the monopolistic case. Thus, while
perfect competition under-performs relative to monopoly in terms of
consumers' anticipatory utility (because media firms fail to incorporate the
data externality), it provides more accurate information, even though it
does not eradicate wrong beliefs due to false narratives.

In Section 5 we return to the baseline model and extend it in two
directions. Our first extension mixes the consumer population with rational
consumers, who know the true model and whose demand for information is
conventionally instrumental. This extension introduces a screening problem
along an unconventional dimension:\ consumers' willingness to adopt a false
hopeful narrative. We show that it is optimal for the media to provide a 
\textit{singleton} menu.\ When there are few rational consumers, this menu
consists of the empowering narrative and a biased signal (albeit less biased
than in the baseline model). When there are few non-rational consumers, the
menu consists of the fatalistic narrative and a fully informative signal.

Next, we consider different separable specifications of the consumer's
utility function. When it is separable in the \textit{state}, the only false
narrative that can outperform the true narrative is the \textit{fatalistic}
narrative. We illustrate this finding with an example in which actions have
objective \textquotedblleft unintended consequences\textquotedblright\ that
the false narrative neglects. When the utility function is separable in the
outcome, the true narrative, coupled with complete information, is optimal.

The extensions cement the main insight of our paper: When media demand is
driven by motivated reasoning, peddling false narratives is a key feature of
media bias.

\section{A Model}

We begin by introducing the primitives of our model. There are four relevant
variables: a state of Nature $t$, an action $a$ taken by a representative
consumer, a signal $s$ that the consumer observes before taking the action
(he can only condition his action on $s$), and an outcome $y$. All variables
take finitely many values. The consumer's vNM utility takes the form:%
\[
u(t,a,y)=v(t,y)-c(a) 
\]%
The objective data-generating process is a probability distribution $p$
defined over the four variables, which can be factorized as follows:%
\begin{equation}
p(t,s,a,y)=p(t)p(s\mid t)p(a\mid s)p(y\mid t,a)  \label{factorization N*}
\end{equation}%
The first and last terms on the R.H.S are exogenously given; they describe
the prior distribution of the state of Nature, and the outcome distribution
conditional on the state and the consumer's action, respectively. Note that
the signal has no direct effect on the outcome. The term $p(s\mid t)$
describes the signal distribution conditional on the state. This
distribution is determined ex-ante by a monopolistic media outlet. Finally,
the term $p(a\mid s)$ represents the consumer's strategy (i.e., his action
distribution conditional on the signal). The strategy's endogenous
determination will be described below.

The causal structure underlying this data-generating process can be
described by the following directed acyclic graph (DAG), denoted $N^{\ast }$:%
\[
\begin{array}{ccc}
t & \rightarrow & s \\ 
\downarrow &  & \downarrow \\ 
y & \leftarrow & a%
\end{array}%
\]%
In this graphical representation, borrowed from the Statistics/AI literature
on probabilistic graphical models (Pearl (2009)), a node represents a
variable, and an arrow represents a direct causal relation. For example, the
link $s\rightarrow a$ means that $s$ is a direct cause of $a$.

Let us now describe the interaction between the media and the representative
consumer. The media moves first, committing ex-ante to a pair $(I,N)$,
where: $I$ is a signal function, which is a Blackwell experiment assigning a
distribution over signals to each state (this is the conditional probability
distribution $(p(s\mid t))_{t,s}$ described above); and $N$ is a narrative,
which is a subset of the two direct causes of $y$.

In particular, the \textit{true narrative} $N^{\ast }$ acknowledges both $t$
and $a$ as the direct causes of $y$. The \textit{\textquotedblleft
empowering\textquotedblright\ narrative} $N^{a}$ postulates that $a$ is the
sole direct cause of $y$. The \textit{\textquotedblleft
fatalistic\textquotedblright\ narrative} $N^{t}$ postulates that $t$ is the
sole direct cause of $y$. Finally, the \textit{\textquotedblleft
denial\textquotedblright\ narrative} $N^{\emptyset }$ postulates that
neither $a$ nor $t$ are direct causes of $y$ (thus implicitly attributing
the outcome to other, unspecified exogenous factors). The three false
narratives ($N^{a},N^{t},N^{\emptyset }$) can be represented by DAGs that
omit at least one of the causal links into $y$ (for example, $N^{a}$ omits
the link $t\rightarrow y$).

Given an objective joint distribution $p$ with full support and the pair $%
(I,N)$, the consumer forms the following conditional belief over $t$ and $y$
given the signal realization $s$ and the action $a$:%
\begin{equation}
\tilde{p}(t,y\mid s,a)=p_{I}(t\mid s)p_{N}(y\mid t,a)
\label{conditional p_N}
\end{equation}%
where $p_{I}(t\mid s)$ is the objective posterior probability of $t$
conditional on $s$, which is induced by the signal function $I$ and given by
Bayes' rule; and $p_{N}(y\mid t,a)$ is the perceived probability of $y$
conditional on $t$ and $a$, which is induced by the narrative $N$.
Specifically,%
\begin{eqnarray*}
p_{N^{\ast }}(y &\mid &t,a)=p(y\mid t,a)\qquad \qquad p_{N^{a}}(y\mid
t,a)=p(y\mid a) \\
p_{N^{t}}(y &\mid &t,a)=p(y\mid t)\qquad \qquad \quad p_{N^{\emptyset
}}(y\mid t,a)=p(y)
\end{eqnarray*}

The interpretation is that $p$ represents a long-run empirical distribution
(reflecting the decisions of other consumers who faced the same problem);
and the narrative $N$ makes sense of this distribution by imposing a
particular explanation for what causes variation in outcomes. The belief $%
p_{N}(y\mid t,a)$ is a systematic, narrative-based distortion of the
objective conditional outcome distribution.

Thus, the media affects the consumer's beliefs via two channels: $(i)$ the
signal function given by $I$, which determines the reporting of current
events; and $(ii)$ the narrative $N$, which provides a perspective --- based
on an interpretation of historical regularities --- for drawing implications
from the signal. The consumer's beliefs about $t$ are induced by $I$,
whereas his beliefs about how $y$ depends on $t$ and $a$ are induced by $N$
(and the long-run objective distribution $p$).

Importantly, when the narrative $N$ is false, $p_{N}(y\mid t,a)$ is not
invariant to the long-run consumer average behavior given by $(p(a\mid
s))_{a,s}$. To see why, elaborate $p_{N}(y\mid t,a)$ for each of the false
narratives:%
\begin{equation}
p_{N^{a}}(y\mid t,a)=\sum_{t^{\prime }}p(t^{\prime }\mid a)p(y\mid t^{\prime
},a)=\sum_{s^{\prime },t^{\prime }}p(s^{\prime }\mid a)p(t^{\prime }\mid
s^{\prime })p(y\mid t^{\prime },a)  \label{p(y|a)}
\end{equation}%
\begin{equation}
p_{N^{t}}(y\mid t,a)=\sum_{a^{\prime }}p(a^{\prime }\mid t)p(y\mid
t,a^{\prime })=\sum_{s^{\prime },a^{\prime }}p(s^{\prime }\mid t)p(a^{\prime
}\mid s^{\prime })p(y\mid t,a^{\prime })  \label{p(y|t)}
\end{equation}%
\begin{eqnarray}
p_{N^{\emptyset }}(y &\mid &t,a)=\sum_{t^{\prime }}p(t^{\prime
})\sum_{a^{\prime }}p(a^{\prime }\mid t^{\prime })p(y\mid t^{\prime
},a^{\prime })  \label{p(y)} \\
&=&\sum_{t^{\prime }}p(t^{\prime })\sum_{s^{\prime },a^{\prime }}p(s^{\prime
}\mid t^{\prime })p(a^{\prime }\mid s^{\prime })p(y\mid t^{\prime
},a^{\prime })\bigskip  \nonumber
\end{eqnarray}%
It is evident that the terms $p(s^{\prime }\mid a)$ and $p(a^{\prime }\mid
s^{\prime })$ involve the consumer's strategy. In other words, long-run
consumer behavior affects narrative-based perception of the mapping from
actions to consequences (given a signal), which in turn affects the
consumer's subjectively optimal decisions. Thus, if we view the long-run
distribution $p$ as a \textit{steady state}, we need an equilibrium notion
of the consumer's subjective optimization.\bigskip

\begin{definition}[Equilibrium]
Given $(I,N)$, a strategy $(p(a\mid s))_{a,s}$ is an $\varepsilon $%
-equilibrium if, whenever $p(a\mid s)>\varepsilon $, $a$ maximizes%
\begin{equation}
V_{I,N}(s,a)=\sum_{t,y}p_{I}(t\mid s)p_{N}(y\mid t,a)u(t,a,y)  \label{V}
\end{equation}%
A strategy is an equilibrium if it is a limit of a sequence of $\varepsilon $%
-equilibria, where $\varepsilon \rightarrow 0$.\bigskip
\end{definition}

This is essentially the definition of personal equilibrium in Spiegler
(2016), and it coincides with Berk-Nash equilibrium (Esponda and Pouzo
(2016)) when the consumer's subjective model is given by the DAG that $N$
induced. The role of trembles in this definition is merely to avoid
conditioning on null events; they play no meaningful role in our analysis.

We assume that the media chooses $(I,N)$ ex-ante to maximize%
\begin{equation}
U(I,N)=\sum_{t}p(t)\sum_{s}p(s\mid t)\max_{a}V_{I,N}(s,a)  \label{U}
\end{equation}%
subject to the constraint that $(p(a\mid s))_{a,s}$ is an equilibrium. The
media's objective function is the consumer's \textit{expected anticipatory
utility}. The idea behind this objective function is that anticipatory
utility drives the consumer's demand for news media. The higher his
anticipatory utility, the greater his media engagement. Our task is to
characterize the media's optimal strategy.

In solving its problem, the media takes into account the consumer's
equilibrium response to the media strategy. This naturally raises the
question of whether the media knowingly anticipates equilibrium effects. One
interpretation is that the media is not aware of them a priori. Instead, it
reacts to past data about consumer engagement, possibly using algorithmic
learning. The equilibrium effects that shape consumers' media engagement
will be reflected in the learning process.\bigskip

\noindent \textit{The necessity of false narratives for media bias}

\noindent Suppose that the media is restricted to providing the true
narrative $N^{\ast }$. This reduces the model to standard information
provision by a sender who can commit ex-ante to a Blackwell experiment. The
sender faces a Bayesian receiver whose indirect utility from a posterior
belief $\mu $ over $t$ is%
\[
\max_{a}\sum_{t}\mu (t)\sum_{y}p(y\mid t,a)u(t,a,y) 
\]%
Since this indirect utility is a maximum over functions that are linear in $%
\mu $, it is convex in $\mu $. Therefore, it is (weakly) optimal for the
sender to commit to a fully informative signal --- i.e., $p(s=t\mid t)=1$
for every $t$. It follows that in our model, given the media's objective of
maximizing the consumer's ex-ante anticipatory utility, the media has no
strict incentive to provide partial or biased information unless it also
peddles a false narrative.

Thus, when the media only supplies information, it treats the consumer as if
he conventionally maximizes his material expected utility. In other words,
it does not matter whether the consumer tries to maximize his actual welfare
or merely his anticipated welfare. Throughout the paper, we refer to the
maximal anticipatory utility attained by the true narrative and complete
information as the \textit{rational-expectations benchmark}.\bigskip

\noindent \textit{Comment on the interpretation of }$a$ and\textit{\ }$y$

\noindent According to one interpretation of our model, $a$ represents a 
\textit{private} action that an individual media consumer takes, and $y$ is
a personal outcome of his choice. For example, $a$ can represent the agent's
career decision or a dietary choice, in which case $y$ represents his
earnings or health outcome, respectively. The data that the consumer relies
on to form beliefs (via the factorization according to $N$) is \textit{%
aggregate}, reflecting the historical choices and outcomes of other
consumers.

An alternative interpretation is that $a$ represents a \textit{public}
choice (such as economic or foreign policy), and $y$ represents a public
outcome (economic growth, national security). According to this
interpretation, the media consumer is a representative \textit{voter}, and
the probability $p(a\mid s)$ is the frequency with which society selects a
political leadership that implements $a$. This is a reduced-form
representation of a democratic process, such that society's choice matches
what the representative voter deems optimal.\bigskip

Our model departs from the canonical information-design framework (see
Bergemann and Morris (2019)), since it allows the designer to influence the
subjective model that the receiver holds. Nevertheless, the assumption that
the consumer always correctly perceives $p(t,s,a)$ ensures that the standard
revelation principle in the information-design literature can be adapted to
the present setting.\bigskip

\begin{remark}[A revelation principle]
Without loss of optimality, we can restrict attention to signal functions
that assign a distribution over recommended actions to each state, and to
equilibria in which $a=s$ with probability one for each $s$.\bigskip
\end{remark}

The proof of this remark follows the footsteps of Theorem 1 in Bergemann and
Morris (2016) --- adapted to the single-player setting --- and is therefore
omitted. The proof involves manipulating the signal function given by $%
(p(s\mid t))_{t,s}$ and the consumer's strategy given by $(p(a\mid s))_{a,s}$%
. In general, when the consumer forms beliefs according to a misspecified
model $N$, such changes may affect $p_{N}(y\mid t,a)$, which could violate
the revelation principle. The reason the principle does hold in our setting
is that the manipulation of $(p(s\mid t))_{t,s}$ and $(p(a\mid s))_{a,s}$ in
the proof leaves $(p(t,a))_{t,a}$ unchanged. As evident from (\ref{p(y|a)})-(%
\ref{p(y)}), this means that $p_{N}(t,a)$ and $p_{N}(y\mid t,a)$ also remain
unchanged, regardless of how $t$ and $a$ are jointly distributed with $s$.
This enables the standard proof to go through. The revelation principle will
simplify our analysis in the sequel.

\section{Analysis}

In this section we analyze the media's optimal strategy. We begin with a
specification that serves as a running example in the paper. We then show
that the qualitative features of the optimal media strategy in our example
are robust.

\subsection{An Example: \textquotedblleft The American
Dream\textquotedblright}

In this example, all variables take values in $\{0,1\}$. The exogenous
components of the data-generating process are:%
\[
p(t=1)=\frac{1}{2}\qquad \qquad p(y=1\mid t,a)=\frac{1}{2}a\left( 2-t\right) 
\]%
The consumer's payoff function is%
\[
u(a,t,y)=ty-ca 
\]%
where $c\in (0,\frac{1}{2})$. (We will later handle the case of $c>\frac{1}{2%
}$.) The action $a$ represents a private decision whether to initiate a
costly economic activity. The outcome $y$ indicates whether the activity is
successful. The state $t$ represents the return from a successful outcome.
High returns are associated with lower chances of a successful outcome.

For a specific example, the consumer is a retail financial investor choosing
whether to engage in active or passive portfolio management. The realization 
$y=1$ represents an increase in the portfolio's value. The state $t$
represents an effective discount factor that reflects economic fundamentals
such as the interest rate for alternative investments, the expected timing
of a successful outcome, or macroeconomic uncertainty. The negative
correlation between $t$ and $y$ has a natural interpretation in this
context. Market fundamentals tend to be reflected in current security
prices. For instance, a low-risk environment is reflected in high current
prices, lowering the prospects of an increase in the portfolio's value.

An alternative story is that the consumer is a college student who decides
how seriously to take his studies. A successful outcome means graduating
(rather than dropping out). The realization $t=1$ represents a college wage
premium. The negative correlation between $t$ and $y$ is due to the fact
that as returns to high education rise, colleges respond by becoming more
demanding, such that graduating becomes less likely.

Under both stories, the media provides information about the fundamentals
represented by $t$, as well as a narrative about what drives the outcome $y$%
. These stories are meant to capture the general idea that media coverage
affects whether a person attains material success or failure is attributed
to either internal factors that the person controls or to external factors
(see Iyengar (1990) for evidence on this in the context of escaping
poverty). Our task is to characterize the optimal media strategy,
considering each of the feasible narratives.\bigskip

\noindent \textit{Rational-expectations benchmark}

\noindent Suppose the media offers the true narrative $N^{\ast }$. As we
saw, it is optimal to couple this narrative with a fully informative signal.
When $t=0$, the consumer knows that $ty=0$, and therefore plays $a=0$. When $%
t=1$, he knows that $p(y=1\mid t=1,a=1)=\frac{1}{2}$. Since $c<\frac{1}{2}$,
the consumer plays $a=1$. It follows that the rational-expectations
benchmark in this example is $\frac{1}{4}-\frac{1}{2}c$.\bigskip

\noindent \textit{Narratives that omit the link }$a\rightarrow y$

\noindent Under the narratives $N^{t}$ and $N^{\emptyset }$, the consumer
believes that his action has no effect on $y$, and therefore prefers to take
the costless action $a=0$. In any equilibrium, $a=0$ with certainty for
every $t$. However, since $y=0$ whenever $a=0$, it follows that $p(y=1\mid
t)=p(y=1)=0$ for every $t$. Therefore, the consumer's anticipatory utility
is necessarily zero, which is below the rational-expectations benchmark. It
follows that the media will necessarily offer a narrative that retains the
link $a\rightarrow y$.\bigskip

\noindent \textit{The empowering narrative}

\noindent Under the narrative $N^{a}$,%
\[
p_{N^{a}}(ty=1\mid a,s)=p(t=1\mid s)p(y=1\mid a) 
\]%
Since $p(y=1\mid a=0)=0$, the consumer's subjective payoff from $a=0$ is
zero for every signal he receives. Let us calculate the consumer's
subjective expected payoff from $a=1$. Denote $q_{t}=p(s=1\mid t)$. By the
revelation principle, we can restrict attention to binary signals and an
equilibrium in which the consumer always plays $a=s$. Then,%
\[
p(y=1\mid a=1)=\frac{1}{2}+\frac{1}{2}p(t=0\mid a=1)=\frac{1}{2}+\frac{1}{2}%
\cdot \frac{q_{0}}{q_{1}+q_{0}} 
\]%
Therefore,%
\begin{equation}
p_{N^{a}}(ty=1\mid s=1,a=1)=\frac{q_{1}}{q_{1}+q_{0}}\cdot \left( \frac{1}{2}%
+\frac{1}{2}\cdot \frac{q_{0}}{q_{1}+q_{0}}\right)  \label{IC s=1}
\end{equation}%
and%
\begin{equation}
p_{N^{a}}(ty=1\mid s=0,a=1)=\frac{1-q_{1}}{2-q_{1}-q_{0}}\cdot \left( \frac{1%
}{2}+\frac{1}{2}\cdot \frac{q_{0}}{q_{1}+q_{0}}\right)  \label{IC s=0}
\end{equation}%
In order for the consumer's strategy to be an equilibrium, we need (\ref{IC
s=1}) and (\ref{IC s=0}) to be weakly above and below $c$, respectively. If
these constraints hold, the consumer's anticipatory utility is $p(s=1)\cdot
\lbrack p_{N^{a}}(ty=1\mid s=1,a=1)-c]$, given by%
\begin{equation}
\frac{q_{1}+q_{0}}{2}\cdot \left[ \frac{q_{1}}{q_{1}+q_{0}}\cdot \left( 
\frac{1}{2}+\frac{1}{2}\cdot \frac{q_{0}}{q_{1}+q_{0}}\right) -c\right]
\label{example payoff}
\end{equation}

Observe that when the media offers a fully informative signal ($q_{1}=1$, $%
q_{0}=0$), this expression coincides with the payoff from $N^{\ast }$. Thus,
if the false narrative $N^{a}$ is to outperform the true narrative, it must
be coupled with incomplete information. We now proceed to calculate the
optimal $I=(q_{0},q_{1})$ that accompanies $N^{a}$.\bigskip

\begin{claim}
\label{claim}When $c<\frac{1}{2}$, it is optimal to set $q_{1}=1$.\bigskip
\end{claim}

Thus, if the optimal signal function has a bias, it must be an optimistic
one, as the media always reports good news ($s=1$) when the state is good ($%
t=1$). The proof of this claim (like all proofs in this paper) is in the
Appendix. The claim implies the following simplified expression for the
consumer's anticipatory utility:%
\begin{equation}
\frac{1}{2}\left[ \frac{1}{2}+\frac{1}{2}\cdot \frac{q_{0}}{1+q_{0}}%
-c(1+q_{0})\right]  \label{simplified payoff}
\end{equation}%
Note that $q_{1}=1$ also implies that (\ref{IC s=0}) is below $c$, such that
playing $a=0$ when $s=0$ is optimal for the consumer. It is now
straightforward to derive the optimal value of $q_{0}$:%
\begin{equation}
q_{0}=\min \left\{ 1,\sqrt{\frac{1}{2c}}-1\right\}  \label{q0}
\end{equation}%
Plugging (\ref{q0}) in (\ref{simplified payoff}), the consumer's ex-ante
anticipatory payoff is%
\[
\begin{array}{ccc}
\frac{1}{2}-\sqrt{\frac{c}{2}} & if & c\in \left[ \frac{1}{8},\frac{1}{2}%
\right) \\ 
\frac{3}{8}-c & if & c\in \left( 0,\frac{1}{8}\right)%
\end{array}%
\]%
which exceeds the rational-expectations benchmark.

Thus, when $c<\frac{1}{2}$, the optimal media strategy involves the
narrative $N^{a}$ coupled with positively biased information: always sending
a good signal in the good state, and sending it with positive probability in
the bad state.

In terms of the interpretation we offered for this example, the false
narrative $N^{a}$ attributes a successful investment outcome entirely to the
consumer's actions, without taking into account the role of the
fundamentals. The accompanying signal function has an optimistic bias in the
direction of claiming that returns are high even when they are not. Thus, on
one hand the media exaggerates the attractiveness of the investment
environment, while on the other hand it suppresses --- via the empowering
narrative --- the negative effect that good fundamentals have on the chances
of good investment outcomes. Thus, we find it apt to refer to the media as
peddling \textquotedblleft the American dream\textquotedblright\ in this
example.

Biased information is necessary for $N^{a}$ to beat the
rational-expectations benchmark. One way to see why is to imagine that the
media provides perfect information. This means that $t$ and $s$ are
perfectly correlated ($s\equiv t$). By the revelation principle, $a$ and $s$
are perfectly correlated ($a\equiv s$). Therefore, $a$ and $t$ are perfectly
correlated. But this means that omitting $t$ as an explanatory variable for $%
y$ does not lead to erroneous beliefs: $p(y\mid a)$ coincides with $p(y\mid
t,a)$. In turn, this means that the consumer effectively has rational
expectations and perfectly monitors $t$, which is precisely the
rational-expectations benchmark. Therefore, imperfect information is
necessary for $N^{a}$ to enhance the consumer's anticipatory utility.

So far, we assumed that $c<\frac{1}{2}$. It is easy to see that when $c\geq 
\frac{1}{2}$, none of the feasible narratives can generate positive utility.
As we already saw, the narratives $N^{t}$ and $N^{\emptyset }$ generate zero
utility for every $c$. It is also immediate from the expressions for the
anticipatory utility induced by $N^{\ast }$ and $N^{a}$ that when $c\geq 
\frac{1}{2}$, these narratives cannot generate positive payoffs.

\subsection{A Characterization Result}

The investment-narrative example has two noteworthy features.\ First, the
empowering narrative emerges as optimal. Second, it is coupled with biased
information that impacts consumer behavior. We now show that both features
hold generally under the model of Section 2.\bigskip

\begin{proposition}
\label{prop Na}If the media can outperform the\ rational-expectations
benchmark, then $N^{a}$ is part of an optimal strategy.\bigskip
\end{proposition}

Thus, the empowering narrative $N^{a}$ is an essential feature of a media
strategy that outperforms the rational-expectations benchmark. The logic
behind the result is as follows. Because $u$ is action-separable, a false
narrative can have an effect on ex-ante anticipatory utility only when it
distorts the joint distribution of $(t,y)$. By definition, the fatalistic
narrative $N^{t}$ cannot do that. In principle, the denial narrative $%
N^{\emptyset }$ can attain such a distortion. However, this effect can be
replicated by $N^{a}$ coupled with no information.

The next result addresses the consumer behavior that the optimal media
strategy induces. We say that the payoff function and the exogenous
data-generating process form a \textit{regular environment} if, under the
true narrative and complete information, the consumer has a unique
best-reply which is a one-to-one function of the state. That is, in regular
environments different states prescribe different unique actions under
rational expectations.\bigskip

\begin{proposition}
\label{prop regular}Suppose the environment is regular. If the optimal media
strategy outperforms the rational-expectations benchmark, then its induced
conditional distribution $(p(a\mid t))_{t,a}$ is different from that
benchmark.\bigskip
\end{proposition}

Thus, when the media deviates from the rational-expectations benchmark, it
necessarily induces changes in consumer behavior. Since regularity assumes a
unique optimal action in each state (under rational expectations), this
means that the outcome induced by the media's strategy departs from what a
paternalistic social planner (aiming to maximize consumers' material
payoffs) would prescribe.

Note that our result does not claim that the media necessarily departs from
fully informative signals. We cannot rule out the possibility that the media
sends a fully informative signal in every state and that the consumer's
subjective best-reply involves mixing, which will be sustained in
equilibrium thanks to the false narrative $N^{a}$.

Regularity plays a key role in the result. To see why, consider the payoff
specification of Section 3.1, and let the data-generating process satisfy $%
p(t=1)=\frac{1}{2}$ and $p(y=1\mid t,a)=1-t$ for every $t,a$. Under rational
expectations, the consumer's optimal action is $a=0$ for every $t$, and the
rational-expectations payoff is $0$ (because $a=0$ and $ty=0$ with
probability one). Using similar arguments as in Section 3.1, it can be shown
that it is optimal for the media to provide $N^{\emptyset }$ (or,
equivalently, $N^{a}$) and no information. The consumer responds by playing $%
a=0$. His anticipatory payoff is $\frac{1}{4}$, beating the
rational-expectations benchmark, although the behavior is the same. Thus,
without regularity, it is possible for the media strategy to outperform the
benchmark without any effect on consumer behavior.

\section{Heterogeneous Consumers}

In this section we extend the model by introducing consumer preference
heterogeneity. Accordingly, the supply side consists of multiple media
strategies that consumers can choose from, each according to his
preferences. We analyze two market structures. In Section 4.1, we consider a
monopolistic media platform acting as a gatekeeper that restricts the entry
of media providers (each represented by a distinct media strategy). The
monopolist's objective is to maximize consumers' aggregate anticipatory
utility --- reflecting the continued assumption that this corresponds to
maximizing their platform engagement. In Section 4.2, we remove the
gatekeeper and analyze a \textquotedblleft perfectly
competitive\textquotedblright\ media market, in which each provider targets
a particular consumer type and tries to maximize his anticipatory utility.

This extension introduces a methodological innovation. While each consumer
type maximizes his own anticipatory utility, this utility --- shaped by the
narrative he adopts --- is evaluated according to the joint distribution
over actions and outcomes, which reflects the \textit{aggregate} behavior of
all consumers. In other words, when consumers adopt a false narrative, they
are subjected to a \textquotedblleft data externality\textquotedblright\
from other consumers. This externality changes the formulation and analysis
of the monopolistic and competitive models of the media market. A key
difference between the two market structures is that the monopolist is an
\textquotedblleft externality maker\textquotedblright\ (who internalizes the
data externality) while competitive media providers are \textquotedblleft
externality takers\textquotedblright . This leads to qualitatively different
characterizations of media strategies that emerge under these market
structures.

\subsection{Monopoly}

In this version of the model, a monopolistic media platform commits ex-ante
to a menu $M$ of pairs $(I,N)$. The set of consumer types is $C=[0,1]$.
Types are distributed according to a continuous and strictly increasing $cdf$
$F$. Let $u_{c}$ be type $c$'s payoff function. Each type $c$ selects a pair 
$(I_{c},N_{c})\in M$ and a signal-dependent action $a_{c}(s)$ to maximize
his ex-ante anticipatory utility. The platform's objective is to maximize
consumers' \textit{aggregate} ex-ante anticipatory utility.

This design problem is a novel type of \textquotedblleft second
degree\textquotedblright\ discrimination, which arises because the platform
cannot prevent consumers from freely choosing their favorite media strategy
from the menu. At first glance, this seems to be a trivial problem, since
there is no conflict of interest between the two parties: Both the consumer
and the monopolist are guided by maximizing consumer anticipatory utility.
However, consumers' choices exert a non-standard externality on one another:
Their narrative-shaped beliefs are all based on the same aggregate
distribution $p$, which reflects the individual choices of all consumers.
This novel interdependence among consumers is what makes the menu-design
problem non-trivial.\bigskip

\noindent \textit{The menu design problem}

\noindent To formally describe the design problem, we begin with how
consumers evaluate alternatives. Fix some profile of consumer types'
media-strategy choices and signal-dependent actions, $%
(I_{c},N_{c},(a_{c}(s)))_{c\in C}.$ Aggregate consumer behavior is given by $%
(p(a\mid t))_{a,t},$ where%
\begin{equation}
p(a\mid t)=\int_{c}\sum_{s}p_{I_{c}}(s\mid t)\cdot 1[a_{c}(s)=a]dF(c)
\label{aggregate consumer}
\end{equation}%
and where $\left( p_{I_{c}}(s\mid t)\right) _{s,t}$ is the Blackwell
experiment given by $I_{c}.$ Denote $\mathbf{a}\equiv (a(s))_{s}$. Given $%
(p(a\mid t))$, consumer type $c$'s ex-ante evaluation of any $(I,N,\mathbf{a}%
)$ is:%
\begin{equation}
U_{c}(I,N,\mathbf{a})=\sum_{s}p_{I}(s)\sum_{t,y}p_{I}(t\mid s)p_{N}(y\mid
t,a(s))u_{c}(t,a(s),y)  \label{U_c}
\end{equation}%
In this formula, $p_{I}(s)$ and $p_{I}(t\mid s)$ are induced by the
objective prior probability $p(t)$ and the Blackwell experiment given by $I$%
. The conditional probability $p_{N}(y\mid t,a)$ is as defined in Section 2,
based on $p(t,a,y)=p(t)p(a\mid t)p(y\mid t,a)$, with $p(a\mid t)$
representing \textit{aggregate} consumer behavior as in (\ref{aggregate
consumer}). Thus, the anticipatory payoff that some type $c$ gets from his
choice of triplet $(I_{c},N_{c},(a_{c}(s)))$ is affected by the choices of
triplets made by all the other types since these determine the joint
aggregate distribution $p(t,a,y).$

In actuality the media platform offers a menu of $(I,N)$ pairs, and consumer
types select items from this menu together with signal-dependent actions. An
optimal menu maximizes consumers' aggregate anticipatory utility such that
consumers' choices and actions satisfy some constraints. An equivalent and
more convenient way to describe the media platform is that it chooses a
profile of \textit{triplets} $(I_{c},N_{c},\mathbf{a}_{c})_{c\in C}$ to
maximize%
\[
\int_{c}U_{c}(I_{c},N_{c},\mathbf{a}_{c})dF(c) 
\]%
subject to the constraints that for every $c\in C$: $(i)$ the triplet $%
(I_{c},N_{c},\mathbf{a}_{c})$ maximizes $U_{c}$ over the set $\{I_{c},N_{c},%
\mathbf{a}_{c}\}_{c\in C}$; and $(ii)$ $\mathbf{a}_{c}$ maximizes $%
U_{c}(I_{c},N_{c},\mathbf{a})$ given $(I_{c},N_{c})$.

To see the \textquotedblleft data externality\textquotedblright\ that the
menu design problem reflects, suppose some consumer types change their
choice of triplet. If this change involves different signal-dependent
actions, it can affect the aggregate distribution $p(a\mid t),$ which in
turn may affect the anticipatory payoff of types who did not change their
choice.\bigskip

\noindent \textit{The \textquotedblleft American dream\textquotedblright\
example revisited}

\noindent Complete characterization of this menu-design problem is beyond
the scope of this paper. Here we make do with applying it to the
\textquotedblleft American dream\textquotedblright\ example of Section 3.1,
extending it by introducing consumer heterogeneity. Specifically, we
identify consumer types with the cost parameter $c$.\footnote{%
Although we only analyze the menu design for this example, we chose to
present the general problem first, because we believe this makes its logic
more transparent.}

In addition, we restrict the domain of feasible information strategies:
Signals are binary, $s\in \{0,1\}$, and the set of feasible signal functions
satisfy $\Pr (s=1\mid t=1)=1$. This restriction entailed no loss of
generality in the representative-consumer case of Section 3.1. This is no
longer the case here; we impose the restriction for tractability, as it
lowers the dimensionality of media strategies. The restriction also means
that we cannot apply the revelation principle. Accordingly, we will not take
it for granted that consumers' actions mimic the signal they receive.

Thus, in what follows, each signal function $I$ is identified with $q$,
which is the probability of submitting $s=1$ when $t=0$. The probability of $%
t=1$ conditional on $s$ is thus%
\begin{equation}
p_{q}(t=1\mid s)=\frac{s}{1+q}  \label{pi}
\end{equation}%
In particular, when the consumer observes the signal $s=0$, he infers that $%
t=0$ and therefore $ty=0$ with probability one. Hence, we can take it for
granted that all consumer types play $a=0$ and earn zero payoffs when
receiving the signal $s=0$.

It follows that $U_{c}(q,N,\mathbf{a})$ can be simplified into%
\begin{eqnarray*}
U_{c}(q,N,\mathbf{a}) &=&p_{q}(s=1)\cdot \lbrack
p_{q}(t=1|s=1)p_{N}(y=1|t=1,a(s=1))-ca(s=1)] \\
&=&\frac{1+q}{2}\cdot \left[ \frac{1}{1+q}p_{N}(y=1|t=1,a(s=1))-ca(s=1)%
\right] \\
&=&\frac{1}{2}p_{N}(y=1|t=1,a(s=1))-\frac{c(1+q)}{2}a(s=1)\bigskip
\end{eqnarray*}%
Likewise, consumers' aggregate state-dependent behavior can be simplified
into%
\[
p(a=1\mid t=1)=\int_{0}^{1}a_{c}(1)dF(c)\qquad \qquad p(a=1\mid
t=0)=\int_{0}^{1}q_{c}a_{c}(1)dF(c) 
\]

We can now restate the platform's problem: Choose a profile $%
(q_{c},N_{c},a_{c}(1))_{c\in C}$ that maximizes%
\[
\int_{0}^{1}U_{c}(q_{c},N_{c},a_{c}(1))dF(c) 
\]%
subject to the constraints that for every $c$, $U_{c}(q_{c},N_{c},a_{c}(1))%
\geq U_{c}(q_{c^{\prime }},N_{c^{\prime }},a_{c^{\prime }}(1))$ for every $%
c^{\prime }\in C$; and that $a_{c}(1)$ maximizes $U_{c}$ given $%
(q_{c},N_{c}) $. The latter constraint can be written as follows:%
\[
\begin{array}{cl}
& \frac{1}{1+q}\cdot p_{N_{c}}(y=1\mid t=1,a=a_{c}(1))-ca_{c}(1) \\ 
\geq & \frac{1}{1+q}\cdot p_{N_{c}}(y=1\mid t=1,a=1-a_{c}(1))-c(1-a_{c}(1))%
\end{array}%
\]%
The rest of this sub-section is devoted to analyzing this menu design
problem.\bigskip

\begin{proposition}
\label{prop general menu}The media maximizes its objective function with a
menu that has the following features:\medskip \newline
$(i)$ The menu includes exactly one pair $(q^{a},N^{a})$; furthermore, $%
q^{a}>0$, and there is $c^{\ast }\in (0,1)$ such that all consumer types in $%
[0,c^{\ast }]$ choose $(q^{a},N^{a})$ and play $a=s$.\medskip \newline
$(ii)$ The menu includes exactly one narrative $N\in \{N^{t},N^{\emptyset
}\} $, coupled with an arbitrary $q$; there is $c^{\ast \ast }\in \lbrack
c^{\ast },1)$ such that all consumer types in $[c^{\ast \ast },1]$ choose $%
(q,N)$ and play $a=0$ with probability one.\medskip \newline
$(iii)$ If $c^{\ast \ast }>c^{\ast }$, then the menu also includes exactly
one pair $(q^{\ast },N^{\ast })$; furthermore, $q^{\ast }<q^{a}$, and all
consumer types in $(c^{\ast },c^{\ast \ast })$ choose $(q^{\ast },N^{\ast })$
and play $a=s$.\bigskip
\end{proposition}

There are a few noteworthy differences from the homogenous case. First,
under homogeneity, a single narrative ($N^{a}$) serves all consumers. The
differentiation between consumer populations (characterized by distinct $c$)
is done through the signal function. In contrast, differentiation between
types in the heterogeneous case is carried out by offering a menu of
narratives. Each of the narratives that keep the link $a\rightarrow y$ is
coupled with a specific signal function. The reason is that given our
restricted domain of signal functions, different media strategies that share
the same narrative are Blackwell-ordered --- and therefore unambiguously
ranked in terms of the anticipatory utility they confer. As a result, no
consumer will select dominated media strategies.

More specifically, the menu includes the narrative $N^{a}$, which is coupled
with biased information toward $a=1$; the narrative $N^{\ast }$ (which need
not be in the menu) has a smaller, potentially zero bias in that direction;
while the other narratives generate the action $a=0$. Thus, we have a
proliferation of narratives, which lead to polarized beliefs and polarized
behavior.

Second, in the homogenous case, market coverage is partial: Consumer types $%
c>\frac{1}{2}$ receive zero payoffs; they are effectively unserved. In
contrast, in the heterogeneous case they earn positive anticipatory payoffs,
thanks to the narratives $N^{t}$ or $N^{\emptyset }$. This is made possible
by the externality between types: Types with high $c$ \textquotedblleft free
ride\textquotedblright\ on low-$c$ types, who play $a=1$ with positive
probability.

The following result completes the characterization of the optimal menu when 
$c$ is uniformly distributed.\bigskip

\begin{proposition}
\label{prop uniform menu}When $c\sim U[0,1]$, the optimal menu consists of
two media strategies: $(q=1,N^{a})$ and $(q=1,N^{\emptyset })$. Consumers
with $c<\frac{3}{11}$ choose the former pair and always play $a=1$; whereas
consumers with $c>\frac{3}{11}$ choose the latter pair and always play $a=0$%
.\bigskip
\end{proposition}

Thus, under uniformly distributed types, the media never provides $any$
information to any consumer. Consumer behavior is highly polarized:
Consumers with high $c$ always play $a=0$ whereas consumers with low $c$
always play $a=1$. What generates this polarization is the different
narratives that the two consumer segments adopt: Low-$c$ consumers opt for
the empowering narrative while high-$c$ consumers opt for the denial
narrative.

\subsection{Perfect Competition}

Let us now consider a competitive media market, in which every media firm is
small and therefore cannot affect aggregate consumer behavior.\bigskip

\begin{definition}[Competitive equilibrium]
A profile $(I_{c},N_{c},\mathbf{a}_{c})_{c\in C}$ is a competitive
equilibrium if for every $c\in C$: $(i)$ $(I_{c},N_{c},a_{c})$ maximizes $%
U_{c}$ over $\{(I_{c},N_{c},\mathbf{a}_{c})\}_{c\in C}$; and $(ii)$ $\mathbf{%
a}_{c}$ maximizes $U_{c}$ given $(I_{c},N_{c})$; where $U_{c}$ is defined as
in (\ref{U_c}).\bigskip
\end{definition}

Unlike the monopoly case, here each media strategy targets a consumer type
and maximizes his anticipatory utility. Media suppliers do not internalize
the data externality between types, because they take the aggregate consumer
behavior implicit in $p$ as given.

As in the previous sub-section, let us apply this definition to the
\textquotedblleft American dream\textquotedblright\ example. As before,
consumer types are identified by their cost parameter. Unlike the previous
sub-section, here we need not restrict the set of feasible signal functions,
except the purely expositional restriction to binary signals that take the
values $0$ or $1$.\bigskip

\begin{proposition}
\label{prop competition}There is an essentially unique competitive
equilibrium. Specifically, there is $\bar{c}\in (0,1)$ given by $2\bar{c}+F(%
\bar{c})=1$, such that: $(i)$ for every $c<\bar{c}$, $I_{c}$ is the fully
informative signal function and $N_{c}=N^{\ast }$; and $(ii)$ for every $c>%
\bar{c}$, $N_{c}=N^{t}$.\bigskip
\end{proposition}

By essential uniqueness, we mean that there could be other media strategies
that implement the same beliefs and actions. For example, when a consumer
chooses $N^{\emptyset }$, the exact signal function is irrelevant for his
beliefs and actions. Also, we could replace $N^{\ast }$ with $N^{a}$ in the
characterization, and consumers' beliefs would be identical.

It follows that for consumer types with low $c$, perfect competition leads
to an unambiguous improvement in the informativeness of news media compared
with monopoly.\ To see why, note that under monopoly, a positive measure of
consumer type with $c$ close to $0$ receive biased information (which is
coupled with the narrative $N^{a}$), whereas under competition they have
rational expectations and full information. When $c\sim U[0,1]$, we have $%
\bar{c}=\frac{1}{3}$, which is above the cutoff $c^{\ast }=\frac{3}{11}$ of
the monopoly case --- i.e., competition improves informativeness for all
consumer types.

\section{Variations}

This section returns to the homogenous-consumer case and explores variations
on the basic model.

\subsection{Introducing Rational Consumers}

So far, we assumed that consumers' demand for news is entirely
non-instrumental --- rather, they use news media to cultivate desirable
beliefs. This gives a role to false narratives as a vehicle for sustaining
such beliefs. Consumers are not dogmatic: They are willing to accept any
narrative, and their sole criterion for selecting a narrative is the
anticipatory utility it induces.

In this sub-section we introduce instrumental motives into the basic model.
There are two ways to do this: assuming that consumers are homogenous but
they mix instrumental and non-instrumental motives; or assuming that the
consumer population consists of conventionally rational types, in addition
to the type described by the basic model. We follow the second route.

Let us return to the basic setting in which $u(t,a,y)=ty-ca$, where all
consumers have the same $c\in (0,\frac{1}{2})$, and introduce a new
heterogeneity. A fraction $\lambda $ of the consumer population have
traditionally instrumental demand for information --- that is, they aim to
maximize their objective expected material payoff, rather than their
anticipatory payoff. Furthermore, these consumers have rational
expectations: They know the true model $N^{\ast }$ and are therefore immune
to narrative peddling by the media. We refer to these consumers as rational,
and to the remaining consumers (who behave as in previous sections) as
non-rational.

The way rational consumers evaluate a media strategy $(I,N)$ is thus quite
simple, because it is equivalent to the way non-rational consumers evaluate
the strategy $(I,N^{\ast })$. The reason is that under the true model $%
N^{\ast }$, the distinction between anticipatory utility and objective
expected material utility disappears: $U(I,N^{\ast })$ is the ex-ante
expected material payoff when the consumer subjectively best-replies to the
beliefs induced by $I$.

We handle the new heterogeneity as we handled the heterogeneity in $c$ in
Section 4. In particular, we adopt the simplifying assumption that $I$ must
involve two signals, $0$ and $1$, such that in state $t=1$ the signal is $%
s=1 $ with probability one. In this setting, this entails no loss of
generality but facilitates exposition by emphasizing the methodological
connection to the model of Section 4.1. Thus, each $I$ is identified with
the probability $q$ of $s=1$ in $t=0$. The media's problem is to choose a
menu of media strategies, $\{(q_{r},N_{r}),(q_{nr},N_{nr})\}$ to maximize%
\[
\lambda \cdot U_{r}(q_{r},N_{r})+(1-\lambda )\cdot U_{nr}(q_{nr},N_{nr}) 
\]%
subject to the constraint that $(q_{i},N_{i})$ maximizes $U_{i}$ for every
type $i\in \{r,nr\}$, given the aggregate distribution induced by the
strategy each type plays; and subject to the constraint that each type $i$'s
strategy is a best-reply given the beliefs induced by the type, the pair $%
(q_{i},N_{i})$ he chooses, and the objective distribution.\bigskip

\begin{proposition}
\label{prop rational}There is an optimal menu that consists of a single pair 
$(q,N)$.\bigskip
\end{proposition}

Thus, the media's problem of screening consumers according to their
rationality is degenerate: The media can offer the same strategy to all
consumers. The result holds more generally when $u(t,a,y)=v(t,y)-ca$, when $%
a,t\in \{0,1\}$, $v$ is an arbitrary function, and $c\in (0,\frac{1}{2})$.

For brevity, we do not provide a detailed derivation of the optimal media
strategy in this environment for all values of $\lambda $. We make do with
illustrating it for extreme values of $\lambda $.\bigskip

\begin{proposition}
\label{prop competition lambda}When $\lambda $ is sufficiently close to $0$,
the optimal media strategy is $(\min \left\{ 1,\sqrt{\frac{1-\lambda }{2c}}%
-1\right\} ,N^{a})$. When $\lambda $ is sufficiently close to $1$, the
optimal media strategy is $(0,N^{t})$.\bigskip
\end{proposition}

Thus, introducing a small group of rational consumers into a population of
non-rational consumers increases the informativeness of the signal that the
media provides, while still keeping the empowering narrative. In contrast,
introducing a small group of non-rational consumers into a population of
rational consumers causes the media to offer the fatalistic narrative while
continuing to give full information.

\subsection{Other Separable Utility Specifications}

In this sub-section we examine alternative specifications of $u(t,a,y)$. All
definitions are adapted straightforwardly.

\subsubsection{An Example: \textquotedblleft Whac-a-Mole\textquotedblright}

Impose the following structure on the data-generating process:%
\[
p(t=1)=\frac{1}{2}\qquad \qquad p(y=1\mid t,a)=\beta (1-a)+(1-\beta
)t\medskip 
\]%
where $\beta \in (\frac{1}{3},1)$. The consumer's payoff function is $%
u(a,t,y)=\mathbf{1}[a=y]$.

We adopt the following interpretation for this specification. The action $a$
represents a public decision how to allocate a scarce resource between two
sectors or locations. For example, the dilemma is whether to allocate
policing effort to one area of criminal activity or another. The state $t$
indicates which sector is more dangerous. The outcome $y$ indicates which
sector ends up being active. Public policy is successful if it allocates the
policing effort to the relevant sector. However, criminal activity exhibits
a \textquotedblleft whac-a-mole\textquotedblright\ property: When the
government cracks down on one area of activity, criminals partly divert
their activity to the other area. This explains the negative correlation
between $a$ and $y$. In this context, the media reports on the dangers posed
by various sectors, and conveys a narrative about what ultimately determines
the active sector. Consumer choice represents support for a certain public
policy (e.g., voting for a political party that runs on this policy).

As before, let us begin our quest for optimal media strategies with the case
in which the narrative is $N^{\ast }$. As usual, we can assume that the
media provides full information. When $t=1$, the consumer's payoff from $a=1$
is $1-\beta $, and the payoff from $a=0$ is $0$. Therefore, the consumer
plays $a=1$ when $t=1$, and his payoff is $1-\beta $. The case of $t=0$ is
handled symmetrically: the consumer plays $a=0$, and earns a payoff of $%
1-\beta $. It follows that the consumer's ex-ante anticipatory utility is $%
1-\beta $. Thus, when the media conveys the true narrative and fully informs
the consumer about $t$, the consumer correctly identifies the dangerous
sector and plays $a=t$. At the same time, the consumer correctly takes the
whac-a-mole effect into account.

We will later see that the narratives $N^{a}$ and $N^{\emptyset }$ (which
omit the link $t\rightarrow y$) are weakly inferior to $N^{\ast }$.
Therefore, let us focus on the narrative $N^{t}$. We apply the revelation
principle and take it for granted that $a=s$ in equilibrium. By definition,%
\[
p_{N^{t}}(y=1\mid s,a)=\sum_{t}p(t\mid s)p(y=1\mid t) 
\]%
where%
\[
p(t=1\mid s=1)=\frac{q_{1}}{q_{0}+q_{1}}\qquad \qquad p(t=1\mid s=0)=\frac{%
1-q_{1}}{2-q_{0}-q_{1}}\medskip 
\]%
\begin{eqnarray*}
p(y &=&1\mid t=1)=\sum_{s}p(s\mid t=1)p(y=1\mid a=s,t=1) \\
&=&q_{1}\cdot (1-\beta )+(1-q_{1})\cdot 1=1-\beta q_{1}
\end{eqnarray*}%
and%
\begin{eqnarray*}
p(y &=&1\mid t=0)=\sum_{s}p(s\mid t=0)p(y=1\mid a=s,t=0) \\
&=&q_{0}\cdot 0+(1-q_{0})\cdot \beta =\beta (1-q_{0})
\end{eqnarray*}%
It follows that the consumer's payoff from playing $a=1$ when $s=1$ is%
\[
U_{N^{t}}(s=1)=\frac{q_{1}}{q_{0}+q_{1}}\cdot (1-\beta q_{1})+\frac{q_{0}}{%
q_{0}+q_{1}}\cdot \beta (1-q_{0}) 
\]%
Likewise, the consumer's payoff from playing $a=0$ when $s=0$ is%
\[
U_{N^{t}}(s=0)=1-\left[ \frac{1-q_{1}}{2-q_{0}-q_{1}}\cdot (1-\beta q_{1})+%
\frac{1-q_{0}}{2-q_{0}-q_{1}}\cdot \beta (1-q_{0})\right] 
\]

In order for this strategy to be an equilibrium, we need both expressions to
be weakly above $\frac{1}{2}$. We will confirm this below. The strategy $a=s$
induces the following ex-ante anticipatory utility:%
\[
\frac{q_{0}+q_{1}}{2}\cdot U_{N^{t}}(s=1)+\left( 1-\frac{q_{0}+q_{1}}{2}%
\right) \cdot U_{N^{t}}(s=0) 
\]%
This expression reduces to%
\[
1+\frac{1}{2}\cdot \left[ (2q_{1}-1)(1-\beta q_{1})-q_{1}\right] +\frac{1}{2}%
\cdot \left[ \beta (2q_{0}-1)(1-q_{0})-q_{0}\right] 
\]%
If the media employs a fully informative signal (i.e., $q_{1}=1$, $q_{0}=0$%
), this expression is equal to $1-\beta $, which is the maximal payoff from
the true narrative $N^{\ast }$. It follows that as in the example of Section
3.1, the false narrative $N^{t}$ can only be optimal when accompanied by
imperfectly informative signals. The optimal signal function is%
\[
q_{1}=\frac{1}{4}+\frac{1}{4\beta }\qquad \qquad q_{0}=\frac{3}{4}-\frac{1}{%
4\beta } 
\]%
Note that the optimal signal treats the two states symmetrically (since $%
q_{0}+q_{1}=1$).

Plugging these values of $q_{0}$ and $q_{1}$, we can confirm that $%
U_{N^{t}}(s)>\frac{1}{2}$ for every $s$. The consumer's ex-ante anticipatory
payoff is%
\[
\frac{(1+\beta )^{2}}{8\beta } 
\]%
which is greater than $1-\beta $.

The false narrative $N^{t}$ that emerges from this exercise neglects the
effect of $a$ on $y$, and thus effectively pretends that the whac-a-mole
effect does not exist. This enables the consumer to be more optimistic about
the success of policies, but only when the narrative is accompanied by
imprecise information.

\subsubsection{A Characterization Result}

The following result shows that the optimality of the narrative $N^{t}$ in
the whac-a-mole example is not a coincidence.\bigskip

\begin{proposition}
\label{prop separable t}Suppose that $u(t,a,y)=v(a,y)+w(t)$. If the media
can outperform the\ rational-expectations benchmark, then $N^{t}$ is part of
an optimal strategy.\bigskip
\end{proposition}

Thus, when $u$ is separable in $t$, the fatalistic narrative is optimal. It
is the analogue of the result that $N^{a}$ is optimal when $u$ is separable
in $a$ (Proposition \ref{prop Na}). It can also be shown that in regular
environments, this narrative can outperform the rational-expectations
benchmark only if it leads to different behavior than the benchmark. The
proof is the same as that of Proposition \ref{prop regular}.

Finally, consider utility functions that are separable in $y$. This turns
out to be a degenerate case, in the sense that it does not give rise to
false narratives.\bigskip

\begin{proposition}
\label{prop separable y}Suppose that $u(t,a,y)=v(t,a)+w(y)$. The benchmark
media strategy is optimal.\bigskip
\end{proposition}

This completes the characterization of utility functions that are separable
in at least one of the variables.\bigskip

\section{Discussion of Related Literature}

This paper belongs to a research program on the role of causal narratives in
economic and political interactions. Eliaz and Spiegler (2020) presented a
modeling framework that formalizes causal narratives as directed acyclic
graphs (building on Spiegler (2016)), where agents' adoption of narratives
is based on the anticipatory utility they generate. Eliaz and Spiegler
(2020) and Eliaz et al. (2022) applied this framework to political
competition. The present paper brings the modeling approach to the market
for news, focusing on the role of media as suppliers of narratives.
Methodologically, its main contributions are: $(i)$ modeling the media's
joint provision of narratives and information; $(ii)$ the novel screening
problem that arises under consumer heterogeneity (in preferences or in
rationality), due to the \textquotedblleft data
externality\textquotedblright\ between consumer types; and $(iii)$ a new
conception of a competitive media market.\footnote{%
Recent empirical and experimental approaches to causal economic narratives
include Ash et al. (2021), Andre et al. (2022), Charles and Kendall (2022),
Macaulay and Song (2023) and Ambuehl and Thysen (2023).}

In terms of economic substance, our paper is part of the literature on media
bias. This phenomenon has been extensively studied from various points of
view. Prat and Str\"{o}mberg (2013) and Gentzkow et al. (2015) provide
comprehensive reviews of this literature. Our paper contributes to a
theoretical strand in this literature that tries to explain media bias as a
demand-based phenomenon arising from non-instrumental aspects of consumers'
attitude to information. The basic idea in this literature is that consumer
derive intrinsic utility from beliefs or from the news they consume,
independently of their effect on decisions. This idea draws on findings in
disciplines outside economics. For example, a meta-study by Hart et al.
(2009) finds that when participants are faced with a choice between
information that supports their prior beliefs and information that may
challenge it, they exhibit a preference for the former. Within the context
of news media, Van der Meer et al. (2020) find evidence that participants
are more likely to view news that confirm their prior beliefs than news that
oppose them.

Mullainathan and Shleifer (2005) attempt to model this phenomenon. They
formalize both states of Nature and news as points along an interval. When a
consumer confronts news, he incurs a cost that increases in the distance
between the news and the mean of his prior belief. Media's strategic choices
are thus reduced to a Hotelling-style model, where the consumer's
psychological cost is analogous to a transportation cost in the standard
Hotelling model.

Gentzkow et al. (2015) present a model in which consumers' utility has two
additively separable components. The first component is a standard material
expected-utility term that employs the consumer's posterior beliefs, which
are obtained conventionally via Bayesian updating. This component treats
beliefs in the usual instrumental manner. The second component is a function
of the consumer's prior belief and the distribution of signals, such that if
the prior leans in the direction of one state, then the function increases
in the frequency of the signal whose label coincides with that state's
label. This captures the idea that people like consuming news that support
their prior beliefs. Note that this non-standard component does not reflect
any belief updating. In particular, if the media always sends a signal that
coincides with the state the consumer deems more likely (such that
effectively the signal is entirely uninformative), the non-instrumental term
reaches its maximal possible level given the consumer's prior belief.

Thus, both Mullainathan and Shleifer (2005) and Gentzkow et al. (2015)
assume that the hedonic effect of news is orthogonal to Bayesian belief
updating. This dissociation between the Bayesian and hedonic aspects of
belief formation is one limitation of existing models we are aware of.
Moreover, even when a consumer behaves as if he has a disutility from a
clash between news and his prior beliefs, this need not be a primitive of
the consumer's preferences, but rather a reflection of the expected change
in his belief as a result of his exposure to news. Thus, it is possible that
what the consumer ultimately cares about his posterior beliefs.

Against this background, our model introduces two innovations. To our
knowledge, it is the first model of news media as suppliers of narratives in
addition to information. It also appears to be the first model in which the
hedonic aspect of media consumers' beliefs is fully integrated with Bayesian
updating of these beliefs. Consumers' intrinsic utility from beliefs is a
function of Bayesian posteriors induced by the information the media
provides and the narrative it peddles. Eliaz and Spiegler (2006) is a
precedent for this aspect of our model. In that paper, we presented of
demand for information --- represented by prior-dependent preferences over
Blackwell experiments --- which is driven by maximization of expected
utility from (correctly specified) Bayesian posterior beliefs. Since that
model allowed for non-convex utility from beliefs, it could accommodate
demand for information that is non-increasing in Blackwell informativeness.
Lipnowski and Mathevet (2018) examined optimal information provision for
agents with such preferences.

Our assumption of Bayesian updating rules out non-Bayesian responses to
information due to motivated reasoning. For example, Taber and Lodge (2006)
show that when subjects are confronted with information that questions their
prior beliefs, they try to discredit it. Thaler (2023) is an experimental
study of the supply of information to agents who exhibit motivated reasoning
(defined as non-Bayesian updating that is affected by the valence of
beliefs). In a similar vein, Furthermore,

The idea that misspecified models can be used to manipulate agents' beliefs
has been studied in other contexts. Eliaz et al. (2021a) analyzed a
cheap-talk model in which the sender provides not only information but also
statistical data (or, equivalently, a model) that enables the receiver to
interpret the information. Eliaz et al. (2021b) characterized the maximal
distortion of perceived correlation between two variables that a causal
model can generate in Gaussian environments. Schwartzstein and Sunderam
(2021) and Aina (2023) studied persuasion problems in which the sender
proposes models, formalized as likelihood functions, and the receiver
chooses among them according to how well they fit historical data. Finally,
our paper is related to a small literature on strategic communication with
agents whose inference from signals departs from the standard Bayesian,
rational-expectations model (e.g., Hagenbach and Koessler (2020), Levy et
al. (2022), de Clippel and Zhang (2022)).

\noindent {\LARGE Appendix: Proofs}\bigskip

\noindent {\large Claim}{\Large \ }{\large \ref{claim}}

\noindent Let us begin by showing that we can set $q_{1}=1$. First, note
that we can rewrite (\ref{example payoff}) as%
\[
\frac{1}{2}\left[ \frac{q_{1}}{2}+\frac{1}{2}\cdot \frac{q_{1}q_{0}}{%
q_{1}+q_{0}}-c(q_{1}+q_{0})\right] 
\]%
The second and third terms inside the brackets are invariant to permuting $%
q_{0}$ and $q_{1}$, whereas the first term is increasing in $q_{1}$ and
invariant to $q_{0}$. Therefore, it is optimal to set $q_{1}\geq q_{0}$.

Second, note that we can rewrite (\ref{example payoff}) as%
\[
\frac{q_{1}+q_{0}}{2}\cdot \left[ \frac{1}{1+\frac{q_{0}}{q_{1}}}\cdot
\left( \frac{1}{2}+\frac{1}{2}\cdot \frac{\frac{q_{0}}{q_{1}}}{1+\frac{q_{0}%
}{q_{1}}}\right) -c\right] 
\]%
Thus, the expression inside the square brackets only depends on the ratio $%
q_{0}/q_{1}$, while the term outside them increases in both $q_{0}$ and $%
q_{1}$. It follows that $q_{1}=1\geq q_{0}$ in optimum. $\blacksquare $%
\bigskip

\noindent {\large Proposition}{\Large \ }{\large \ref{prop Na}}

\noindent Denote $\min_{a}c(a)=c^{\ast }$. Consider the narrative $N^{t}$.
In this case, the consumer believes that $a$ has no causal effect on $y$.
Therefore, for every $s$, he will only mix over actions that minimize $c$.
Then, in equilibrium,%
\[
\begin{array}{ccc}
U(I,N^{t}) & = & \sum_{s}p(s)\sum_{a}p(a\mid s)\sum_{t}p(t\mid
s)\sum_{y}p(y\mid t)v(t,y)-c^{\ast } \\ 
& = & \sum_{s}p(s)\sum_{t}p(t\mid s)\sum_{y}\left( \sum_{a^{\prime
}}p(a^{\prime }\mid t)p(y\mid t,a^{\prime })\right) v(t,y)-c^{\ast } \\ 
& = & \sum_{t}p(t)\sum_{s}p(s\mid t)\sum_{y}\left( \sum_{a^{\prime
}}p(a^{\prime }\mid t)p(y\mid t,a^{\prime })\right) v(t,y)-c^{\ast } \\ 
& = & \sum_{t}p(t)\sum_{a^{\prime }}p(a^{\prime }\mid t)\sum_{y}p(y\mid
t,a^{\prime })v(t,y)-c^{\ast }%
\end{array}%
\]%
Since $c(a^{\prime })=c^{\ast }$ whenever $p(a^{\prime }\mid t)>0$, the last
expression can be rewritten as%
\[
\sum_{t}p(t)\sum_{a^{\prime }}p(a^{\prime }\mid t)\left[ \sum_{y}p(y\mid
t,a^{\prime })v(t,y)-c(a^{\prime })\right] 
\]%
which is by definition weakly below%
\[
\sum_{t}p(t)\max_{a}\left[ \sum_{y}p(y\mid t,a)v(t,y)-c(a)\right] 
\]%
The final expression is the rational-expectations benchmark. Therefore, $%
N^{t}$ cannot be part of a media strategy that outperforms it.

Now consider the narrative $N^{\emptyset }$. As in the previous case, the
consumer believes that $a$ has no effect on $y$. Therefore, for every $s$, $%
p(a\mid s)>0$ only if $a$ minimizes $c$. It follows that the consumer's
ex-ante anticipatory utility is%
\[
\begin{array}{ccl}
U(I,N^{\emptyset }) & = & \sum_{s}p(s)\sum_{t}p(t\mid
s)\sum_{y}p(y)v(t,y)-c^{\ast } \\ 
& = & \sum_{t}p(t)\sum_{y}p(y)v(t,y)-c^{\ast } \\ 
& = & \sum_{t}p(t)\sum_{y}\left( \sum_{a^{\prime }}p(a^{\prime })p(y\mid
a^{\prime })\right) v(t,y)-c^{\ast } \\ 
& = & \sum_{t}p(t)\sum_{a^{\prime }}p(a^{\prime })\sum_{y}\left[ p(y\mid
a^{\prime })v(t,y)-c(a^{\prime })\right] \\ 
& \leq & \sum_{t}p(t)\max_{a}\left[ \sum_{y}p(y\mid a)v(t,y)-c(a)\right]%
\end{array}%
\]%
The final expression is the ex-ante anticipatory utility induced by the
narrative $N^{a}$ coupled with no information. It follows that the maximal
anticipatory utility from $N^{\emptyset }$ can be replicated by the
narrative $N^{a}$ (coupled with fully uninformative signals). $\blacksquare $%
\bigskip

\noindent {\large Proposition}{\Large \ }{\large \ref{prop regular}}

\noindent Assume the contrary --- i.e., suppose there is a media strategy
that induces the same $(p(a\mid t))_{t,a}$ as in the rational-expectations
benchmark, yet outperforms it.

We first show that $N^{a}$ is the only narrative that can be part of the
strategy. The proof of Proposition \ref{prop Na} showed that $N^{t}$ can
never outperform the benchmark; and $N^{\ast }$ cannot do so by definition.
Now consider $N^{\emptyset }$. Under this narrative, the consumer will
assign probability one to $\arg \min_{a}c(a)$ for every $t$. By assumption,
this is also the consumer's behavior under rational expectations, but this
contradicts the definition of regularity. This leaves $N^{a}$ as the only
possible narrative.

By regularity, $p(a\mid t)$ assigns probability one to a distinct action for
each $t$. Let $t(a)$ be the unique state for which $a$ is played under $p$.
Since $t=t(a)$ whenever $p(t,s,a)>0$, it follows that%
\[
p(y\mid a)=p(y\mid t,a)
\]%
for every $(t,a)$ in the support of $p$. Consequently,%
\[
p_{N^{a}}(t,y\mid s,a)=p(t,y\mid s,a)
\]%
and therefore, the consumer's anticipatory utility under $p$ and $N^{a}$ is
equal to the rational-expectations benchmark, a contradiction. $\blacksquare 
$\pagebreak 

\noindent {\large Proposition}{\Large \ }{\large \ref{prop general menu}}

\noindent The proof proceeds stepwise.\bigskip

\noindent \textbf{Step 1}: \textit{Without loss of generality, each
narrative is coupled with a unique }$q$.$\smallskip $

\noindent Assume the contrary --- i.e., $M$ contains two pairs $(q,N)$ and $%
(q^{\prime },N)$ with $q^{\prime }<q$. This means that the signal function
given by $q^{\prime }$ Blackwell-dominates the signal function given by $q$
(recall that $\Pr (s=1\mid t=1)=1$ under both functions). Any consumer type $%
c$ who compares the two pairs will weakly prefer $(q^{\prime },N)$. The
reason is that consumers take the objective distribution $p$ as given. Since
both pairs share the same narrative $N$, they both induce the same $%
p_{N}(y\mid t,a)$. This reduces the comparison between the pairs to a
standard comparison between signal functions by an expected-utility
maximizer.

Consider consumer types $c$ who choose $(q,N)$ from $M$. They must be
indifferent between this pair and $(q^{\prime },N)$. Except for a
zero-measure set of types, this can only be the case if the consumers take a
constant action given each of the pairs (if their subjectively optimal
action were state-contingent, then $q$ and $q^{\prime }$ would induce
different ex-ante expected utility). Moreover, this must be the same
constant action since otherwise, only a particular (zero measure) type would
be ex-ante indifferent between $(q,N)$ and $(q^{\prime },N)$. Now suppose we
remove $(q,N)$ from the menu. Then, since $(q^{\prime },N)$ was optimal for
these consumer types under $M$, this will continue to be the case and all
these consumers will therefore choose $(q^{\prime },N)$. This will have no
effect on the aggregate consumer strategy because $(q^{\prime },N)$ and $%
(q,N)$ induce the same choices by consumers who chose $(q,N)$ from $M$.
Therefore, the switch by these consumers from $(q,N)$ to $(q^{\prime },N)$
has no effect on how other consumer types evaluate any media strategy. It
follows that without loss of generality, we can remove $(q,N)$ from the
menu. $\square $\bigskip

\noindent \textbf{Step 2}: \textit{Under the optimal menu, a positive
measure of consumer types play }$a=1$\textit{\ with positive probability.}$%
\smallskip $

\noindent Assume that under the optimal menu, all consumer types play $a=0$
with certainty. Then, regardless of the media strategy they choose, their
anticipatory utility is $0$. This is obviously the case for consumer types
who choose $N^{\ast }$ or $N^{a}$, because these narratives induce the
correct belief that $a=0$ causes $y=0$ with certainty.

As to types who choose $N^{t}$, they estimate the conditional probability 
\[
p(y=1\mid t)=\sum_{a}p(a=1\mid t)\cdot \frac{2-t}{2}=0 
\]%
for every $t$. Therefore, these types earn zero anticipatory utility as well.

Finally, types who choose $N^{\emptyset }$ form the correct belief that $%
p(y=1)=0$ (because $a=0$ with probability one by assumption, and $p(y=1\mid
a=0)=0$). It follows that all types earn zero anticipatory utility. However,
if the monopolist offers the singleton menu consisting of the media strategy 
$(0,N^{\ast })$, every type $c<\frac{1}{2}$ will earn $\frac{1}{4}-\frac{c}{2%
}>0$, a contradiction.\bigskip

\noindent \textbf{Step 3}: \textit{Interval structure of types' choices}$%
\smallskip $

\noindent By Step 1, we can assume that for each feasible narrative $N$
there is at most one $q$ such that $(q,N)\in M$. Because the narratives $%
N^{t}$ and $N^{\emptyset }$ omit $a$ as a cause of $y$, any consumer who
chooses a media strategy that includes one of these narratives will always
play $a=0$. Furthermore, whenever a consumer chooses a media strategy that
includes $N^{\ast }$ or $N^{a}$, he will play $a=0$ in response to $s=0$,
since $t=0$ with certainty conditional on $s=0$. Therefore, by Step 2, $M$
must include a pair $(q,N)$ such that $N\in \{N^{\ast },N^{a}\}$, and there
is a positive measure of consumer types who select this pair and play $a=1$
in response to $s=1$.\medskip

\noindent \textit{How consumer types rank }$(q^{\ast },N^{\ast })$\textit{\
and }$(q^{a},N^{a})$\textit{\ when playing }$a=s$\textit{\ in response to
both pairs}

\noindent Suppose that $M$ includes both $(q^{\ast },N^{\ast })$ and $%
(q^{a},N^{a})$ such that for each of these pairs, there is a positive
measure of consumer types who choose it and play $a=1$ in response to $s=1$.
The ex-ante anticipatory utility that these pairs induce for a consumer of
type $c$ is:%
\begin{eqnarray*}
U_{c}(q^{\ast },N^{\ast }) &=&\frac{1}{2}p(y=1\mid t=1,a=1)-\frac{1+q^{\ast }%
}{2}c\medskip \\
&=&\frac{1}{2}\cdot \frac{1}{2}[2-1]-\frac{1+q^{\ast }}{2}c=\frac{1}{4}-%
\frac{1+q^{\ast }}{2}c
\end{eqnarray*}%
and%
\begin{eqnarray*}
U_{c}(q^{a},N^{a}) &=&\frac{1}{2}p(y=1\mid a=1)-\frac{1+q^{a}}{2}c\medskip \\
&=&\frac{1}{2}\cdot \frac{1}{2}[2-p(t=1\mid a=1)]-\frac{1+q^{a}}{2}c
\end{eqnarray*}%
It is immediate that $U_{c}(q^{\ast },N^{\ast })>U_{c}(q^{a},N^{a})$ only if 
$q^{\ast }<q^{a}$. Therefore, $q^{a}>0$ in this case. Consequently, if $%
U_{c}(q^{\ast },N^{\ast })>U_{c}(q^{a},N^{a})$, then $U_{c^{\prime
}}(q^{\ast },N^{\ast })>U_{c^{\prime }}(q^{a},N^{a})$ for every $c^{\prime
}>c$. It follows that if both $(q^{\ast },N^{\ast })$ and $(q^{a},N^{a})$
are in $M$ and induce $a=1$ in response to $s=1$, then the set of types who
choose $(q^{\ast },N^{\ast })$ lies above the set of types who choose $%
(q^{a},N^{a})$.\medskip

\noindent \textit{Showing that }$M$\textit{\ includes }$(q^{a},N^{a})$%
\textit{\ without loss of generality}

\noindent Suppose that $M$ does not include $(q^{a},N^{a})$. Then, $M$
includes a pair $(q^{\ast },N^{\ast })$ such that a positive measure of
consumer types choose this pair and play $a=1$ in response to $s=1$. (The
reason is that consumers always play $a=0$ in response to $N^{t}$ or $%
N^{\emptyset }$, as well as in response to $N^{\ast }$ when $s=0$.) Now add $%
(q^{\ast },N^{a})$ to the menu. It is evident that $U_{c}(q^{\ast
},N^{a})\geq U_{c}(q^{\ast },N^{\ast })$ for every $c$. Therefore, if
playing $a=s$ is optimal given $(q^{\ast },N^{\ast })$, then it is also
optimal given $(q^{\ast },N^{a})$.

Moreover, if types who previously chose $(q^{\ast },N^{\ast })$ and played $%
a=s$ switch to $(q^{\ast },N^{a})$ and thus continue to play $a=s$, this
switch does not change the joint aggregate distribution $p(t,a)$ because $%
(q^{\ast },N^{a})$ and $(q^{\ast },N^{\ast })$ share the same signal
function and induce the same consumer strategy.

Finally, consider types who previously chose a media strategy that induces $%
a=0$ for all $s$ now switch to $(q^{\ast },N^{a})$. By revealed preferences,
the switch improves their own anticipatory utility, hence they must play $%
a=s $ (because if they play $a=0$, their anticipatory utility is $0$). At
the same time, the switch does not affect $p(t=1\mid a=1)$ because this
probability is equal to 
\[
\frac{\frac{1}{2}\left( m(q^{\ast },N^{a})+m(q^{\ast },N^{\ast })\right) }{%
\frac{1}{2}\left( m(q^{\ast },N^{a})+m(q^{\ast },N^{\ast })\right) +\frac{1}{%
2}q^{\ast }\left( m(q^{\ast },N^{a})+m(q^{\ast },N^{\ast })\right) }=\frac{1%
}{1+q^{\ast }} 
\]%
where $m(q^{\ast },N^{a})+m(q^{\ast },N^{\ast })$ is the total mass of types
who choose either $(q^{\ast },N^{\ast })$ or $(q^{\ast },N^{a})$ (which is
precisely the mass of types who choose $a=s$). As a result, the switch does
not affect $U_{c}(q^{\ast },N^{a})$ for any $c$. By definition, it also does
not affect $U_{c}(q^{\ast },N^{\ast })$.

It follows that we can sustain an equilibrium with weakly higher aggregate
anticipatory utility when $(q^{\ast },N^{a})$ is added to the menu. Thus,
the menu will contain some pair $(q^{a},N^{a})$ that induces $a=1$ in
response to $s=1$.\medskip

\noindent \textit{The set of types who choose }$(q^{a},N^{a})$\textit{\ and
play }$a=s$

\noindent We now establish that there is $c^{\ast }>0$ such that all types
in $[0,c^{\ast })$ choose $(q^{a},N^{a})$ and play $a=s$. To see why,
suppose first that $(q^{\ast },N^{\ast })$ is in $M$ and that there is a
positive measure of consumers who select this pair and play $a=s$. Then, as
we showed, the set of types who select $(q^{a},N^{a})$ over $(q^{\ast
},N^{\ast })$ and play $a=s$ lies to the left of the set of types who choose 
$(q^{\ast },N^{\ast })$ and play $a=s$.

Now suppose $(q^{a},N^{a})$ is the only media strategy in $M$ that induces $%
a=s$. If type $c$ prefers $(q^{a},N^{a})$ to a media strategy that induces
him to always play $a=0$, then so does every $c^{\prime }<c$.

Thus, the set of types who prefer $(q^{a},N^{a})$ and play $a=1$ in response
to $s=1$ is at the low end of $[0,1]$.\medskip

\noindent \textit{The set of types who always play }$a=0$

\noindent Suppose first that $M$ includes $(q^{\ast },N^{\ast })$ such that
a positive measure of types choose this pair and play $a=s$. The payoff from
this choice is $\frac{1}{4}-\frac{1}{2}c$, which is negative for $c>\frac{1}{%
2}$. Thus, such types will respond to the same pair by always playing $a=0$.

Now suppose that the only media strategy that induces $a=1$ with positive
probability is $(q^{a},N^{a})$. But then,%
\[
U_{c}(q^{a},N^{a})=\frac{1}{2}\cdot \frac{1}{2}[2-\frac{1}{1+q^{a}}]-\frac{%
1+q^{a}}{2}c 
\]%
which is negative for $c\approx 1$. Hence, such types will respond to $%
(q^{a},N^{a})$ by always playing $a=0$.

Clearly, every type $c$ who chooses $N^{t}$ or $N^{\emptyset }$ always plays 
$a=0$.

Finally, if type $c$ prefers a media strategy that induces him to always
play $a=0$, then so does type $c^{\prime }>c$. Therefore, there must be $%
c^{\ast \ast }<1$ such that all types $c>c^{\ast \ast }$ choose a media
strategy that induces $a=0$ for all $s=0$. \medskip

To conclude this step, there are two cutoffs, $0<c^{\ast }\leq c^{\ast \ast
}<1$, such that: $(i)$ all types below $c^{\ast }$ choose $(q^{a},N^{a})$
and play $a=s$; $(ii)$ all types between $c^{\ast }$ and $c^{\ast \ast }$
choose $(q^{\ast },N^{\ast })$ and play $a=s$; and $(iii)$ all types above $%
c^{\ast \ast }$ play $a=0$ with probability one. $\square $\bigskip

\noindent \textbf{Step 4}: \textit{The menu contains exactly one of the
narratives }$N^{t}$\textit{\ or }$N^{\emptyset }\smallskip $

\noindent First, observe that the menu need not include both $N^{t}$ and $%
N^{\emptyset }$. The reason is that both narratives induce $a=0$ with
probability one, such that they only potentially differ in the subjective
ex-ante expected value of $ty$ that they induce. In particular, they exert
the same externality on other types. Therefore, the menu will include only
the one that yields the higher payoff.

Second, suppose that $M$ contains neither $N^{t}$ nor $N^{\emptyset }$.
Then, types above $c^{\ast \ast }$ will select the narratives $N^{\ast }$ or 
$N^{a}$ and always play $a=0$, thus obtaining a payoff of $0$. Suppose we
add $N^{t}$ or $N^{\emptyset }$ coupled with no information. By Step 2, $a=1$
with positive probability, and therefore, both narratives will induce
strictly positive payoff for types above $c^{\ast \ast }$. We need to
examine the possibility that lower types will switch from $(q^{a},N^{a})$ or 
$(q^{\ast },N^{\ast })$ to the new media strategy. However, if a type $%
c<c^{\ast \ast }$ deviates in this direction, then so does every $c^{\prime
}\in (c,c^{\ast \ast })$. Consider two cases.\medskip

\noindent \textit{Case }$1$: $N^{\ast }$ is not in $M$. In this case, the
deviation does not change $p(t=1\mid a=1)$ because this quantity is not
affected by increasing the share of consumers who always play $a=0$, and the
set of consumers who play $a=s$ all induce the same $\Pr (a\mid t)$.
Therefore, it does not affect $U_{c}(q^{a},N^{a})$ for any $c$. By revealed
preference, the deviation improves the ex-ante payoff of the deviating
types. It follows that there is an unambiguous increase in aggregate
consumer payoffs.\medskip

\noindent \textit{Case }$1$: $N^{\ast }$ is in $M$. In this case, the set of
deviating types is some interval $[c^{\ast \ast \ast },c^{\ast \ast }]$. As
a result, since $q^{\ast }<q^{a}$, this deviation lowers $p(t=1\mid a=1)$
and therefore increases $U_{c}(q^{a},N^{a})$ for any $c$. Moreover, it has
no effect on $U_{c}(q^{\ast },N^{\ast })$ by definition. By revealed
preference, the deviation improves the ex-ante payoff of the deviating
types. It follows that the deviation increases aggregate consumer payoffs,
even after taking into account the equilibrium effects of this deviation due
to the data externality. $\square \bigskip $

It remains to show that $q^{a}>0$ if $c^{\ast }=c^{\ast \ast }$ --- i.e.,
the only narrative that induces $a=1$ with positive probability is $N^{a}$.
We know from the homogenous case that for every $c<\frac{1}{2}$, $q^{a}>0$
attains higher utility than $q^{a}=0$ (and recall that the utility from $%
q^{a}=0$ for $c>\frac{1}{2}$ cannot be positive). Moreover, since $q^{a}>0$
generates a higher overall probability of $a=1$, it exerts a positive
externality than $q=0$ on consumers who choose $N^{\emptyset }$ (it has no
externality on consumers who choose $N^{t}$). Therefore, deviating from $%
q^{a}=0$ to $q^{a}>0$ is strictly profitable for the media. $\blacksquare $%
\bigskip

\noindent {\large Proposition}{\Large \ }{\large \ref{prop uniform menu}}

\noindent The proof proceeds stepwise, taking the characterization in
Proposition \ref{prop general menu} as a starting point.\bigskip

\noindent \textbf{Step 1}: $c^{\ast }=c^{\ast \ast }\smallskip $

\noindent Assume that $c^{\ast \ast }>c^{\ast }$. The payoffs induced by $%
(q^{\ast },N^{\ast })$ and $(q^{a},N^{a})$ at some $c$ are%
\begin{eqnarray*}
U_{c}(q^{\ast },N^{\ast }) &=&\frac{1}{4}-\frac{1+q^{\ast }}{2}c \\
U_{a}(q^{a},N^{a}) &=&\frac{1}{4}\left[ 2-p(t=1\mid a=1)\right] -\frac{%
1+q^{a}}{2}c
\end{eqnarray*}%
In the proof of Step 3 of Proposition \ref{prop general menu}, we showed
that $q^{a}>q^{\ast }$. Since $c\sim U[0,1]$, we can write%
\[
p(t=1\mid a=1)=\frac{c^{\ast \ast }}{c^{\ast \ast }+c^{\ast }q^{a}+(c^{\ast
\ast }-c^{\ast })q^{\ast }}=\frac{c^{\ast \ast }}{c^{\ast \ast }(1+q^{\ast
})+c^{\ast }(q^{a}-q^{\ast })} 
\]%
At $c^{\ast }$, the indifference between $(q^{\ast },N^{\ast })$ and $%
(q^{a},N^{a})$ can be written as follows:%
\[
\frac{1}{2}c^{\ast }(q^{a}-q^{\ast })=\frac{1}{4}\left[ 1-\frac{c^{\ast \ast
}}{c^{\ast \ast }(1+q^{\ast })+c^{\ast }(q^{a}-q^{\ast })}\right] \medskip 
\]

Observe that if we slightly raise $c^{\ast }$ and lower $q^{a}$ such that $%
q^{a}$ is still above $q^{\ast }$ and $c^{\ast }(q^{a}-q^{\ast })$ remains
unchanged, then the indifference condition continues to hold, as long as we
keep $c^{\ast \ast }$ fixed. In this way, $p(t=1\mid a=1)$ remains
unchanged. This modified consumer strategy is an equilibrium and it is
strictly profitable for the media. To see why, note first that $c^{\ast \ast
}$ is unchanged because by construction, $p(a=1)$ and $p(a=1\mid t=1)$ are
both unchanged, hence the payoff from $N^{t}$ or $N^{\emptyset }$ is
unchanged. Since the payoff from $(q^{\ast },N^{\ast })$ is by definition
invariant to $(p(a\mid t))$, the indifference at $c^{\ast \ast }$ continues
to hold. Thus, the set of types who always play $a=0$ and their utility are
unaffected. Now consider the infra-marginal types $c<c^{\ast }$. These types
are now better off thanks to the decrease in $q^{a}$, and since $p(a=1\mid
t=1)$ is unchanged. The types who chose and continue to choose $(q^{\ast
},N^{\ast })$ are unaffected by definition. Therefore, the new equilibrium
is an improvement, a contradiction.

What this step establishes is that we can restrict attention to menus $M$
and consumer strategies that take either of the two following forms:\newline
$(i)$ $M=\{\{q^{a},N^{a}),(q^{t},N^{t})\}$, all consumer types in $%
[0,c^{\ast }]$ choose $(q^{a},N^{a})$ and play $a=s$, and all consumer types 
$c>c^{\ast }$ choose $(q^{t},N^{t})$ and play $a=0$; and\newline
$(ii)$ $M=\{\{q^{a},N^{a}),(q^{\emptyset },N^{\emptyset })\}$, all consumer
types in $[0,c^{\ast }]$ choose $(q^{a},N^{a})$ and play $a=s$, and all
consumer types $c>c^{\ast }$ choose $(q^{\emptyset },N^{\emptyset })$ and
play $a=0$. $\square $\bigskip

\noindent \textbf{Step 2}: \textit{Completing the characterization when }$M$%
\textit{\ includes }$N^{t}\smallskip $

\noindent Aggregate utility under $M=\{\{q^{a},N^{a}),(q^{t},N^{t})\}$ is%
\[
\int_{0}^{c^{\ast }}U_{c}(q^{a},N^{a})dc+\int_{c^{\ast
}}^{1}U_{c}(q^{t},N^{t})dc 
\]%
where%
\[
U_{c}(q^{a},N^{a})=\frac{1}{4}\left[ 2-\frac{1}{1+q^{a}}\right] -\frac{%
1+q^{a}}{2}c 
\]%
and%
\begin{eqnarray*}
U_{c}(q^{t},N^{t}) &=&p(ty=1)=p(t=1)\cdot p(y=1\mid t=1)\medskip \\
&=&\frac{1}{2}\cdot p(a=1\mid t=1)\cdot \frac{1}{2}(2-1)=\frac{1}{4}c^{\ast }
\end{eqnarray*}%
Thus, the objective function can be written as%
\begin{eqnarray*}
&&\int_{0}^{c^{\ast }}\left\{ \frac{1}{4}\left[ 2-\frac{1}{1+q^{a}}\right] -%
\frac{1+q^{a}}{2}c\right\} dc+(1-c^{\ast })\cdot \frac{1}{4}c^{\ast }\medskip
\\
&=&c^{\ast }\cdot \frac{1}{4}\left[ 2-\frac{1}{1+q^{a}}\right] -\frac{1+q^{a}%
}{2}\cdot \frac{1}{2}(c^{\ast })^{2}+(1-c^{\ast })\cdot \frac{1}{4}c^{\ast }
\end{eqnarray*}

The cutoff $c^{\ast }$ satisfies%
\[
\frac{1}{4}\left[ 2-\frac{1}{1+q^{a}}\right] -\frac{1+q^{a}}{2}c^{\ast }=%
\frac{1}{4}c^{\ast } 
\]%
Plugging this equation into the objective function, we obtain%
\[
\frac{2q^{a}+1}{\left( 2q^{a}+3\right) ^{2}} 
\]%
The optimal value of $q^{a}$ is $\frac{1}{2}$, yielding an aggregate utility
of $\frac{1}{8}$. $\square $\bigskip

\noindent \textbf{Step 3}: \textit{Completing the characterization when }$M$%
\textit{\ includes }$N^{\emptyset }\smallskip $

\noindent Aggregate utility under $M=\{\{q^{a},N^{a}),(q^{\emptyset
},N^{\emptyset })\}$ is%
\[
\int_{0}^{c^{\ast }}U_{c}(q^{a},N^{a})dc+\int_{c^{\ast
}}^{1}U_{c}(q^{\emptyset },N^{\emptyset })dc 
\]%
where%
\[
U_{c}(q^{a},N^{a})=\frac{1}{4}\left[ 2-\frac{1}{1+q^{a}}\right] -\frac{%
1+q^{a}}{2}c 
\]%
and%
\begin{eqnarray*}
U_{c}(q^{\emptyset },N^{\emptyset }) &=&p(t=1)\cdot p(y=1) \\
&=&p(t=1)\cdot \lbrack p(t=1)\cdot p(y=1\mid t=1)+p(t=0)\cdot p(y=1\mid t=0)]
\\
&=&\frac{1}{2}\cdot \lbrack \frac{1}{2}\cdot p(a=1\mid t=1)\cdot \frac{1}{2}%
(2-1)+\frac{1}{2}\cdot p(a=1\mid t=0)\cdot \frac{1}{2}(2-0)] \\
&=&\frac{1}{2}\cdot \lbrack \frac{1}{2}\cdot c^{\ast }\cdot \frac{1}{2}(2-1)+%
\frac{1}{2}\cdot c^{\ast }q^{a}\cdot \frac{1}{2}(2-0)] \\
&=&\frac{1}{2}\cdot \lbrack \frac{1}{4}c^{\ast }+\frac{1}{2}c^{\ast }q^{a}]
\\
&=&\frac{c^{\ast }}{4}[\frac{1}{2}+q^{a}]
\end{eqnarray*}%
Thus, the objective function can be written as%
\begin{eqnarray*}
&&\int_{0}^{c^{\ast }}\left\{ \frac{1}{4}\left[ 2-\frac{1}{1+q^{a}}\right] -%
\frac{1+q^{a}}{2}c\right\} dc+(1-c^{\ast })\cdot \frac{c^{\ast }}{4}[\frac{1%
}{2}+q^{a}] \\
&=&c^{\ast }\cdot \frac{1}{4}\left[ 2-\frac{1}{1+q^{a}}\right] -\frac{1+q^{a}%
}{2}\cdot \frac{1}{2}(c^{\ast })^{2}+(1-c^{\ast })\cdot \frac{c^{\ast }}{4}[%
\frac{1}{2}+q^{a}]
\end{eqnarray*}

The cutoff $c^{\ast }$ satisfies%
\[
\frac{1}{4}\left[ 2-\frac{1}{1+q^{a}}\right] -\frac{1+q^{a}}{2}c^{\ast }=%
\frac{c^{\ast }}{4}[\frac{1}{2}+q^{a}] 
\]%
Plugging this equation into the objective function, we obtain%
\[
\frac{3}{4}\left( 2q^{a}+1\right) ^{2}\frac{2q^{a}+3}{\left( 6q^{a}+5\right)
^{2}\left( q^{a}+1\right) } 
\]%
This expression is monotonically increasing in $q^{a}$, hence the optimal
value of $q^{a}$ is $1$, yielding an aggregate utility of approximately $%
0.139$. $\square $\bigskip

Since the menu characterized by Step 3 yields a higher payoff than the one
characterized by Step 2, the optimal menu includes the denial narrative, and
sets $q^{a}=1$. $\blacksquare $\bigskip

\noindent {\large Proposition}{\Large \ }{\large \ref{prop competition}}

\noindent First, we establish that without loss of generality, $I_{c}$ is
the perfectly informative signal function for every $c$. The reason is that
the maximization of type $c$'s anticipatory utility takes $p$ as given
without taking into account the effect of the behavior induced by $%
(I_{c},N_{c})$ on $p_{N}$. Therefore, $U_{c}$ is effectively the maximum of
functions that are linear in beliefs, hence it is convex in posterior
beliefs. It follows that a fully informative signal maximizes $U_{c}$ (as in
the rational-expectations benchmark). It is the unique maximizer if it
induces $a=s$.

Second, we show that all consumers play $a=0$ when $t=0$. This holds under $%
N^{t}$ or $N^{\emptyset }$ because under these narratives, $a$ has no causal
effect on $y$. Under $N^{a}$ or $N^{\ast }$, optimal provision of
information implies that when $t=0$ the consumer knows that $ty=0$, and
therefore finds $a=0$ optimal.

An immediate consequence of the previous step is that $p(t=0\mid a=1)=0$.
such that the formulas for $U_{c}$ under $N^{\ast }$ and $N^{a}$ coincide.
Thus, from now on, we will take it for granted that the only narrative that
can induce $a=1$ with positive probability is $N^{\ast }$. Let us denote by $%
\sigma $ the fraction of consumers who play $a=1$ when $t=1$.

Third, we will show that $N^{t}$ weakly outperforms $N^{\emptyset }$ for
every consumer type. To see why, let us write down the anticipatory utility
under each of these narratives. The anticipatory utility under $N^{t}$ is%
\[
p(t=1)p(y=1\mid t=1)=\frac{1}{2}\cdot \sigma \cdot \frac{1}{2}(2-1)=\frac{1}{%
4}\sigma 
\]%
The anticipatory utility under $N^{\emptyset }$ is%
\[
\begin{array}{ccc}
p(t=1)p(y=1) & = & \frac{1}{2}\cdot \left[ \frac{1}{2}\cdot p(y=1\mid t=1)+%
\frac{1}{2}\cdot p(y=1\mid t=0)\right] \\ 
& = & \frac{1}{4}p(y=1\mid t=1) \\ 
& = & \frac{1}{8}\sigma%
\end{array}%
\]%
Note that $p(y=1\mid t=0)=0$ because all consumers play $a=0$ when $t=0$.

Thus, the only narratives we need to consider are $N^{\ast }$ and $N^{t}$.
Moreover, we can assume that any consumer who adopts $N^{\ast }$ will play $%
a=1$ when $t=1$, because otherwise he would get zero payoffs, which is below
the payoff he can get from $N^{t}$. A consumer of type $c$ will prefer $%
N^{\ast }$ if $\frac{1}{4}-\frac{c}{2}>\frac{1}{4}\sigma $. There is a
cutoff $\bar{c}$ characterized by $\frac{1}{4}-\frac{\bar{c}}{2}=\frac{1}{4}%
F(\bar{c})$, such that all $c<\bar{c}$ choose $N^{\ast }$ and play $a=t$,
while all $c>\bar{c}$ choose $N^{t}$ and always play $a=0$. $\blacksquare $%
\bigskip

\noindent {\large Proposition}{\Large \ }{\large \ref{prop rational}}

\noindent Suppose first $N_{nr}\in \{N^{t},N^{\emptyset }\}$. Then, type $nr$%
's evaluation of the pair $(q_{nr},N_{nr})$ is independent of $q_{nr}$, and
he always plays $a=0$ in response to this pair. As to type $r$'s evaluation
of the same pair, it is independent of $N_{nr}$ because this type applies $%
N^{\ast }$. Thus, type $r$'s evaluation of $(q_{nr},N_{nr})$ only depends on 
$q_{nr}$. Moreover, it is decreasing in the degree of media bias $q$. It
follows that if the media collapses the menu $%
\{(q_{r},N_{r}),(q_{nr},N_{nr})\}$ into a singleton $\{(0,N_{nr})\}$, it
will weakly raise consumers' aggregate anticipatory utility.

Now suppose $N_{nr}=N^{a}$. First, it cannot be the case that type $nr$
responds to $(q_{nr},N_{nr})$ by always playing $a=0$. The reason is that
under this strategy, $U_{nr}(q_{nr},N^{a})=0$. This cannot be optimal,
because the singleton menu $\{(0,N^{t})\}$ would outperform it: Since type $%
r $ responds to this pair by playing $a=t$, $U_{nr}(0,N^{t})=\frac{1}{4}%
\lambda >0$ --- i.e., the singleton menu maximizes $U_{r}$ while inducing $%
U_{nr}>0$.

We can thus take it for granted that type $nr$ plays $a=s$. If $%
(q_{r},N_{r})\neq (q_{nr},N_{nr})$, then in order for type $r$ to favor $%
(q_{r},N_{r})$ over $(q_{nr},N_{nr})$, he must find the former pair more
informative --- i.e., $q_{r}<q_{nr}$. Therefore, $q_{r}<1$, such that type $%
r $ plays $a=s$ in response to $(q_{r},N_{r})$.

Then,%
\[
U_{r}(q_{r},N_{r})=\frac{1}{4}-\frac{1+q_{r}}{2}c
\]%
and%
\begin{eqnarray*}
U_{r}(q_{nr},N^{a}) &=&\frac{1}{4}\left[ 2-p(t=1\mid a=1)\right] -\frac{%
1+q_{nr}}{2}c \\
&=&\frac{1}{4}\left[ 2-\frac{p(t=1)p(a=1\mid t=1)}{p(t=1)p(a=1\mid
t=1)+p(t=0)p(a=1\mid t=0)}\right]  \\
&&-\frac{1+q_{nr}}{2}c \\
&=&\frac{1}{4}\left[ 2-\frac{1}{1+\lambda q_{r}+(1-\lambda )q_{nr}}\right] -%
\frac{1+q_{nr}}{2}c
\end{eqnarray*}%
Crucially, $p(t=1\mid a=1)$ is based on the aggregate distribution, which
makes use of both types' strategies. Denote $\bar{q}=\lambda
q_{r}+(1-\lambda )q_{nr}$. It follows that the aggregate anticipatory
utility is%
\begin{equation}
\lambda \frac{1}{4}+(1-\lambda )\frac{1}{4}\left[ 2-\frac{1}{1+\bar{q}}%
\right] -\frac{1+\bar{q}}{2}c  \label{N^a utility mixed}
\end{equation}%
This is exactly the same aggregate utility that would be obtained from a
singleton menu $\{(\bar{q},N^{a})\}$, where both types respond to the pair
by playing $a=s$. therefore, there is no loss of generality in restricting
ourselves to singleton menus. $\blacksquare $\pagebreak 

\noindent {\large Proposition}{\Large \ }{\large \ref{prop competition
lambda}}

\noindent Suppose $\lambda $ is close to $0$ --- i.e., the population
consists almost entirely of non-rational consumers. Then, $N^{a}$ is part of
an optimal strategy. This follows from continuity relative to the $\lambda
=0 $ case. The derivative of (\ref{N^a utility mixed}) with respect to $\bar{%
q}$ is%
\begin{equation}
\frac{1-\lambda }{4(1+\bar{q})^{2}}-\frac{c}{2}  \label{derivative N^a}
\end{equation}%
The optimal $\bar{q}$ is given by the first-order condition.

Now suppose $\lambda $ close to $1$ --- i.e., the population consists almost
entirely of rational consumers. Suppose that $N^{a}$ is part of an optimal
strategy. As $\lambda \rightarrow 1$, (\ref{derivative N^a}) becomes
negative, hence it is optimal to set $\bar{q}=0$ --- i.e., offering a fully
informative signal. However, when $\bar{q}=0$, $N^{a}$ offers the same
utility for non-rational consumers as $N^{\ast }$, hence it cannot
outperform $(0,N^{\ast })$. The only remaining media strategies we need to
check are $(0,N^{t})$ and $(0,N^{\emptyset })$. Note that 
\[
U_{nr}(0,N^{t})=\frac{1}{4}\lambda >\frac{1}{8}\lambda
=U_{nr}(0,N^{\emptyset }) 
\]%
Obviously, since $\lambda \approx 1$, $U_{nr}(0,N^{t})>U_{r}(0,N^{t})=\frac{1%
}{4}-\frac{1}{2}c$. It follows that the optimal media strategy when $\lambda 
$ is close to $1$ is $(0,N^{t})$. $\blacksquare $\bigskip

\noindent {\large Proposition}{\Large \ }{\large \ref{prop separable t}}

\noindent First, observe that for every feasible strategy $(I,N)$, the
ex-ante subjective expectation of $w(t)$ is 
\[
\sum_{s}p(s)\sum_{t^{\prime }}p_{N}(t^{\prime }\mid s)w(t^{\prime }) 
\]%
Recall that for every feasible narrative $N$, $p_{N}(t^{\prime }\mid
s)\equiv p(t^{\prime }\mid s)$. Therefore, the above expression reduces to 
\[
\sum_{t^{\prime }}p(t^{\prime })w(t^{\prime })=Ew(t) 
\]%
regardless of $(I,N)$. Therefore, we can regard $Ew(t)$ as a constant in the
media's objective function, and focus on the $v$ term. Thus, from now on, we
conveniently set $w(t)=0$ for all $t$ --- this without loss of generality.

Consider the narrative $N^{a}$. In this case,%
\[
U_{I,N^{a}}(s,a)=\sum_{t}p(t\mid s)\sum_{y}p(y\mid a)v(a,y)=\sum_{y}p(y\mid
a)v(a,y)
\]%
We can see that $I$ is irrelevant for the consumer's anticipatory utility
from action $a$. It follows that his ex-ante anticipatory utility can be
written as

\[
\begin{array}{ccc}
\sum_{a}p(a)\sum_{y}p(y\mid a)v(a,y) & = & \sum_{a}p(a)\sum_{y}\left(
\sum_{t}p(t\mid a)p(y\mid t,a)\right) v(a,y) \\ 
& = & \sum_{t}p(t)\sum_{a}p(a\mid t)\sum_{y}p(y\mid t,a)v(a,y) \\ 
& \leq  & \sum_{t}p(t)\max_{a}\sum_{y}p(y\mid t,a)v(a,y)%
\end{array}%
\]

The final expression is the consumer's maximal ex-ante anticipatory utility
according to the true narrative $N^{\ast }$. Therefore, $N^{a}$ cannot be
part of a media strategy that outperforms the strategy of providing complete
information and the true narrative.

Now consider the narrative $N^{\emptyset }$. In this case,%
\[
U_{I,N^{\emptyset }}(s,a)=\sum_{t}p(t\mid
s)\sum_{y}p(y)v(a,y)=\sum_{y}p(y)v(a,y) 
\]%
Here, too, we can see that $I$ is irrelevant for the consumer's anticipatory
utility from action $a$. It follows that his ex-ante anticipatory utility
can be written as%
\begin{eqnarray*}
&&\sum_{a}p(a)\sum_{y}p(y)v(a,y)=\sum_{a}p(a)\sum_{y}\left(
\sum_{t}p(t)p(y\mid t)\right) v(a,y) \\
&=&\sum_{a}p(a)\sum_{t}p(t)\sum_{y}p_{N^{t}}(y\mid t,a)v(a,y)
\end{eqnarray*}%
This is equal to the ex-ante anticipatory utility from the mixture over
actions $(p(a))$, when the media conveys the narrative $N^{t}$ and provides
no information. It follows that the maximal anticipatory utility from $%
N^{\emptyset }$ can be replicated by the narrative $N^{t}$ (coupled with
fully uninformative signals). $\blacksquare $\bigskip

\noindent {\large Proposition}{\Large \ }{\large \ref{prop separable y}}

\noindent Consider the term $v(t,a)$. As we have observed, $%
p_{N}(t,s,a)\equiv p(t,s,a)$ for every feasible narrative $N$. Therefore,%
\[
\sum_{s}p(s)\sum_{a}p(a\mid s)E_{N}v(t,a\mid s,a)=E_{N^{\ast }}(v(t,a)) 
\]%
Now turn to the term $w(y)$. The ex-ante expectation of this term according
to some feasible $(I,N)$ is%
\[
\sum_{y}p_{N}(y)w(y) 
\]%
We will now show that $p_{N}(y)\equiv p_{N^{\ast }}(y)$ for every feasible
false narrative. First, observe that%
\[
p_{N}(y)=\sum_{t}p(t)\sum_{s}p(s\mid t)\sum_{a}p(a\mid s)p_{N}(y\mid
t,a)=\sum_{t,a}p(t,a)p_{N}(y\mid t,a) 
\]%
Let us now write this expression for each of the three feasible false
narratives. For $N^{a}$,%
\[
\sum_{t,a}p(t,a)p(y\mid a)=\sum_{a}p(a)p(y\mid a)=p(y) 
\]%
For $N^{t}$,%
\[
\sum_{t}p(t,a)p(y\mid t)=\sum_{t}p(t)p(y\mid t)=p(y) 
\]%
Finally, for $N^{\emptyset }$,%
\[
\sum_{t,a}p(t,a)p(y)=p(y) 
\]%
It follows that both terms of $u$ are undistorted by any false narrative.
Therefore, the media cannot outperform the true narrative (coupled with full
information). $\blacksquare $

\end{document}